\documentclass[11pt,dvips,twoside,letterpaper]{article} 
\usepackage[dvipdfm]{graphicx}
\usepackage{amsmath}
\usepackage{bm}
\usepackage{fancyhdr}
\usepackage{makeidx}


\makeatletter
\def\cleardoublepage{\clearpage\if@twoside \ifodd\c@page\else%
    \hbox{}%
    \thispagestyle{empty}%
    \newpage%
    \if@twocolumn\hbox{}\newpage\fi\fi\fi}
\makeatother

\renewcommand{\thesection}{\arabic{section}.}
\renewcommand{\thesubsection}{\thesection\arabic{subsection}.}

\def\figurename{Figure}
\makeatletter
\renewcommand{\fnum@figure}[1]{\figurename~\thefigure.}
\makeatother

\def\tablename{Table}
\makeatletter
\renewcommand{\fnum@table}[1]{\tablename~\thetable.}
\makeatother
\renewcommand{\headrulewidth}{0pt}
\newcommand\beq{\begin{equation}}
\newcommand\eeq{\end{equation}}
\newcommand\beqa{\begin{eqnarray}}
\newcommand\eeqa{\end{eqnarray}}

\makeatletter
\def\ps@fancy{%
\def\chaptermark##1{\markboth{\ifnum \c@secnumdepth>\z@
Chapter \thechapter\hskip 1em\relax \fi ##1}{}}%
\def\sectionmark##1{\markright{\ifnum \c@secnumdepth>\z@
\thesection\hskip 1em\relax \fi ##1}{}}%
\def\subsectionmark##1{\markright {\ifnum \c@secnumdepth >\@ne
\thesubsection\hskip 1em\relax \fi ##1}}%
\ps@@fancy
\gdef\ps@fancy{\@fancyplainfalse\ps@@fancy}%
\ifdim\headwidth<0sp
\global\advance\headwidth123456789sp\global\advance\headwidth\textwidth
\fi
}

\newcommand{\be}{\begin{eqnarray}}
\newcommand{\ee}{\end{eqnarray}}

\begin{document}
\title{
{\begin{flushleft}
\vskip 0.45in
{\normalsize\bfseries\textit{Chapter~}}
\end{flushleft}
\vskip 0.45in
\bfseries\scshape Thermodynamical description of hadron-quark phase transition
 and its implications on compact-star phenomena}}
\author{\bfseries\itshape Nobutoshi Yasutake$^1$\thanks{E-mail address: nobutoshi.yasutake@it-chiba.ac.jp}, Tsuneo Noda$^2$,
Hajime Sotani$^3$,\\
\bfseries\itshape Toshiki Maruyama$^4$,
and Toshitaka Tatsumi$^5$\\
$^1$Department of Physics, Chiba Institute of Technology,\\ 2-1-1 Shibazono,Narashino, Chiba 275-0023, Japan\\
$^2$Department of Physics, Kyushu University,\\ 6-10-1 Hakozaki, Higashi-ku, Fukuoka, 812-8581 Japan\\
$^3$Division of Theoretical Astronomy,\\ National Astronomical Observatory of Japan,\\
2-21-1 Osawa, Mitaka, Tokyo 181-8588, Japan\\
$^4$Advanced Science Research Center,\\ Japan Atomic Energy Agency,\\
Shirakata Shirane 2-4, Tokai, Ibaraki 319-1195, Japan\\
$^{5}$Department of Physics, Kyoto University,\\ Kyoto 606-8502, Japan.
}
\date{}
\maketitle
\thispagestyle{empty}
\setcounter{page}{1}









\

\begin{abstract}
One of the most promising possibilities may be the appearance of quark matter in
astrophysical phenomena in the light of recent progress in observations.
The mechanism of deconfinement is not well understood,
but the thermodynamical aspects of the hadron-quark (HQ) phase transition have been
extensively studied in recent years.
Then the mixed phase of hadron and quark matter becomes important;
the proper treatment is needed to describe the HQ phase transition
and derive the equation of state (EOS) for the HQ matter,
based on the Gibbs conditions for phase equilibrium.
We here adopt a EOS based on the baryon-baryon interactions including hyperons for the hadron phase, while we use rather simple EOS within the MIT bag model for the quark phase.
One of the interesting consequences
may be the appearance of the inhomogeneous
structures called ``pasta'', which are brought about by the surface and the Coulomb interaction effects.
We present here a comprehensive review of our recent works
about the HQ phase transition in various astrophysical situations: cold catalyzed matter,
hot matter and neutrino-trapped matter.
We show how the pasta structure becomes unstable by
the charge screening of the Coulomb interaction, thermal effect or the neutrino trapping effect.
Such inhomogeneous  structure may affect astrophysical phenomena
through its elasticity or thermal properties.
Here we also discuss some implications on supernova explosion, gravitational wave and cooling of compact stars.
\end{abstract}

\pagestyle{fancy}
\fancyhead{}
\fancyhead[EC]{Nobutoshi Yasutake,Tsuneo Noda  et al.}
\fancyhead[EL,OR]{\thepage}
\fancyhead[OC]{Thermodynamical description of hadron-quark ...}
\fancyfoot{}
\renewcommand\headrulewidth{0.5pt}

\section{Introduction}

The mechanism of the hadron-quark~(HQ) deconfinement is one of hot topics in modern physics. Theoretically, numerical simulations based on the lattice gauge theory have been done to estimate the critical temperature and figure out properties of the phase transition\cite{petreczky12}. However, at present, these attempts do not provide definite results for the deconfinement transition in high density region. 
On the other hand, there have been done many theoretical studies by using the effective models of QCD\cite{QCD, alf08}. Therefore, one of the most hopeful strategies to get insight into  this mechanism is to find or predict some signals about QCD phase transition in the astrophysical phenomena such as supernovae, black hole formations, mergers of neutron stars, cooling of compact stars, etc., in the light of recent theoretical and observational progress.
Then the equation of state~(EOS) is a basic element to discuss the implications on above astrophysical phenomena.
There have been proposed and discussed various types of scenario concerned with the HQ deconfinement transitions at high-density region, but
it is still unclear, even, whether the phase transition is the cross over or the first order~\cite{fukushima10}. Only assumption in this review is that the phase transition is
of the first order as suggested by many model studies \cite{QCD}.
One of the direct consequences of this assumption is the emergence of the hadron-quark (HQ) mixed phase during the phase transition.
In this review we first discuss the thermodynamical aspects of the HQ transition, and present some implications on astrophysical phenomena subsequently.

Historically, the mixed phases have been often treated very naively by applying the Maxwell construction to get the EOS in thermodynamic equilibrium.
In 1990 Glendenning has pointed out that the phase equilibrium in multi-component system or in the case of plural chemical potential corresponding to the conserved quantities such as in neutron star matter must be considered based on the more general Gibbs conditions \cite{glen, gle92}. He emphasized that the neutron-star matter should be treated as a{\it  binary} system specified by two conserved quantities, electromagnetic charge and baryon numbers.
Thus the pressure becomes no more constant as in the Maxwell construction, but still increases in the mixed phase. He also demonstrated that the mixed phase spreads in a rather wide density region within a bulk calculation, where two kinds of semi-infinite matter are considered in phase equilibrium, and the pressure balance, chemical equilibrium and the global charge-neutrality conditions are imposed, discarding the Coulomb interaction.
It is important to see that particle fractions in each phase are no more the same for given average particle-fractions. Generally any phase transition in a system with two or more conserved charges must be {\it non-congruent}
\footnote{When one dare to apply the Maxwell construction in such systems, one must violate the Gibbs conditions by ,e.g., imposing local charge neutrality.}
.
However, the Coulomb interaction is discarded in this calculation and we can by no means estimate how important it is, or the surface energy can not be taken into account a priori.
If we consider a more realistic situation, we must consider inhomogeneous matter with various geometric structures called ``{\it pasta}'', taking into account the finite-size effects such as the Coulomb interaction and the surface tension. The pasta resembles the {\it nuclear pasta} in liquid-gas transition at subnuclear densities~\cite{Lorenz1993, Oyamatsu1993, SOT1995, maru2005, new, oka}.
Heiselberg et al.\ have demonstrated that such finite-size effects are important for the mixed phase in the context of the HQ transition \cite{HPS93}.
Treating the pasta structure as in ref.~\cite{Rav83}, they simply considered uniform particle densities in both phases. They showed a possibility of that the finite-size effects may largely restrict the region of the mixed phase.
Subsequently, Voskresensky et al.\ have shown that the rearrangement of particle densities and the screening effect for the Coulomb interaction are important, and  the large surface tension gives the mechanical instability of the geometrical structures in the pasta phase \cite{voskre}. Using the effective model for hadron matter and the MIT bag model for quark matter,
Endo et al.\  explicitly demonstrated these features \cite{end05}. Following this idea, we have studied the pasta structures brought by the kaon condensation and the HQ deconfinement transition \cite{marukaon06,end05,hyp07}. More recently we have extended our framework to deal with the finite temperature case or the neutrino-trapping matter, which may be relevant to supernova explosions or neutron star mergers \cite{yas09, yas12}.

In this review we first present thermodynamic concepts of the non-congruent transition of the HQ phase transition (Section \ref{cong}). In the section we will also present some  results about the phase transition, where the effects of finite temperature or neutrino trapping are also discussed. In Section \ref{astro} is discussed some astrophysical implications of the HQ phase transition. Section \ref{sum} is devoted to summary and concluding remarks.

\section{Phase Equilibrium in Thermodynamics on Hadron-Quark Phase Transition}
\label{cong}

In this section we discuss the emergence of the pasta structure associated
with the HQ deconfinement transition\footnote{
We, hereafter, dismiss the possibility of color superconductivity \cite{ARRW}.}.
For our purpose we need both EOS's of hadron and quark matter as realistically as possible.
As we mentioned in the last section, no one knows how to exactly calculate the HQ phase transition at high density region.
The studies by using the effective models of QCD such as the MIT bag model or the NJL model have been actively done instead.
We, hereafter, use the MIT bag model for simplicity.

We adopt a realistic EOS for the hadron matter based on the microscopic nucleon-nucleon~($NN$), and nucleon-hyperon~($NY$) interactions.
Although there are many models to describe the EOS with hyperons,
they have many uncertainties which come from the uncertainty of $NY$ and $YY$ interactions. However, nowadays, the potentials of the baryon-baryon interactions have been directly calculated by the lattice QCD calculations~\cite{nemura09}. We hope that they provide realistic $NY$ and $YY$ interactions in near future. 
Motivated by such background, it becomes a very hot topic to calculate the EOS from the baryon-baryon interaction directly. Here, we show the EOS by the theoretical framework of the nonrelativistic Brueckner-Hartree-Fock~(BHF) approach~\cite{baldo98a} based on the microscopic $NN$ and $NY$ interactions. The BHF calculation is a reliable and well-controlled theoretical approach for the study of dense baryon matter.

\begin{figure}[hbt]
\begin{center}
\includegraphics[width=1\textwidth]{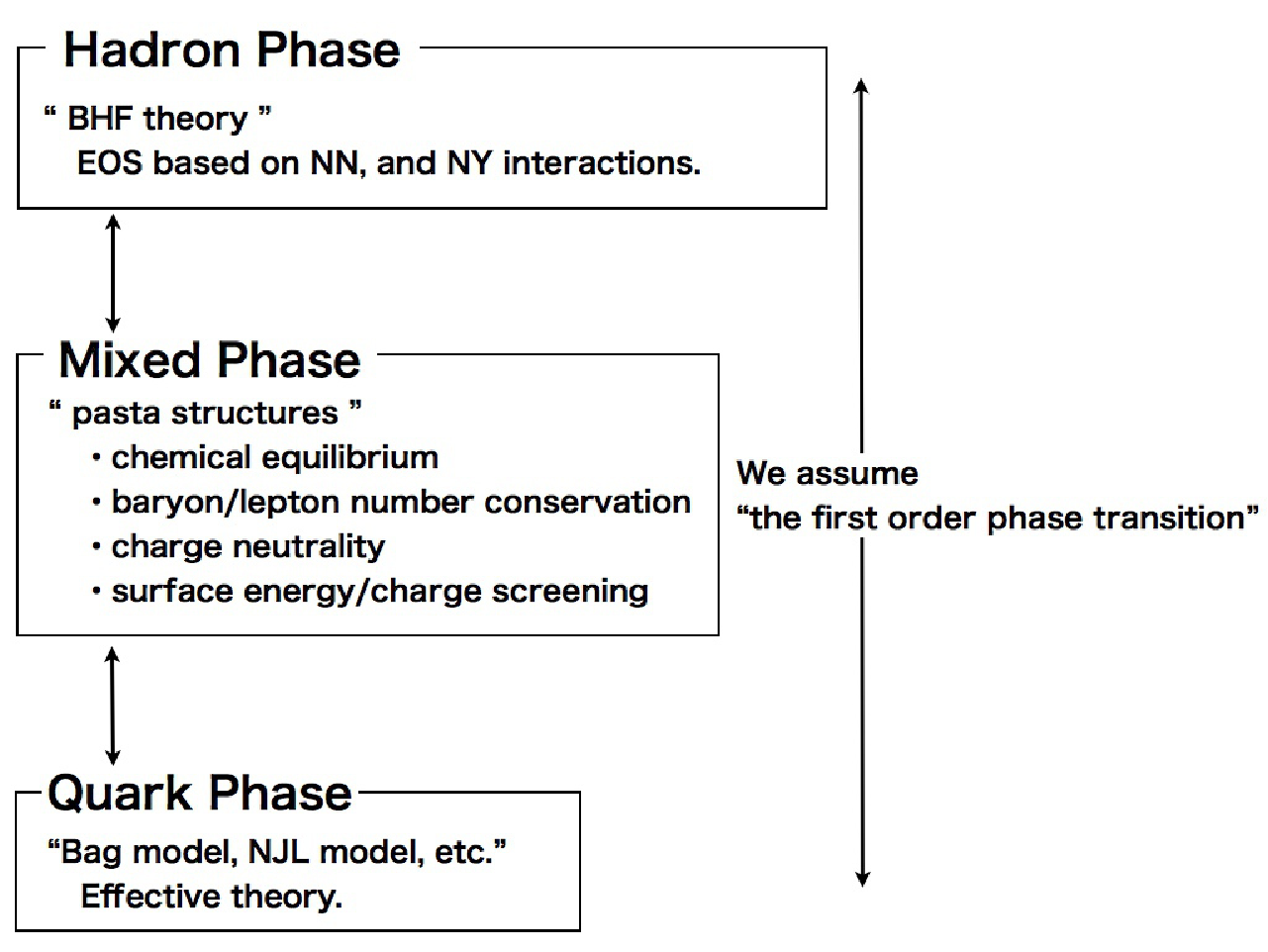}
\end{center}
\caption{Schematic picture of our strategy on the study about the HQ phase transition for the astrophysical implications.}
\label{fig:strt}
\end{figure}

In Fig.~\ref{fig:strt}, we sum up our strategy to construct EOS including the HQ phase for the astrophysical implications.
Since our goal is to figure out the realistic properties of the HQ phase transition,
we chose one of the most realistic ways to know the HQ mixed phase at high-density region; $NY$, $YY$ interactions will be clear in the future for the hadron phase.

Then we consider the phase-equilibrium by including the finite-size effects. The Coulomb interaction and the surface energy are taking into account in a consistent way:
charge screening effect is automatically took into account by solving the Poisson equation, but the surface energy at the interface of hadron and quark matter is
treated in a phenomenological way by using the surface tension parameter.
Thus, the largest ambiguity in our study is the uncertainty in the quark models.
Hence, one of the main purposes of our study is to avoid this ambiguity as much as possible by comparing our results with astrophysical phenomena. Here we sum up our method, and report our current status.

\subsection{Hadron Phase; the Brueckner-Hartree-Fock theory}

BHF theory is well-known  to describe hadron matter using the realistic baryon-baryon interaction.
In this review we do not explain the details of the procedure, but they can be found in Refs.\ \cite{baldo98a,baldo98b,schulze95,vidana95}.
We adopt the Argonne $V_{18}$ potential~\cite{wiringa95} for $NN$ interaction,
and semi-phenomenological Urbana UIX nucleonic three-body forces~\cite{pudliner97}
and the Nijmegen soft-core NSC89 $NY$ potentials \cite{maessen89}.
Unfortunately, there are not reliable potentials for the hyperon-hyperon interaction now.
Therefore, we neglect them here.

To take into account the effects of finite temperature, we adopt \textit{Frozen Correlations Approximation}~\cite{baldo99, nicotra06a,nicotra06b}.
The correlations at finite temperature are assumed to be the same with the ones at zero temperature.
It is found to be a good approximation at finite temperature
as Refs.~\cite{baldo99, nicotra06a,nicotra06b}.

In the following, we briefly describe our framework.
First, we get the chemical potential $\mu_i$  from the number density $n_i$,
\begin{eqnarray}
n_i     &=& \cfrac{g}{(2\pi)^3} \int^\infty_0 f_i(p) ~ 4 \pi p^2dp     \\
f_i(p) &=& \cfrac{1}{ {\rm exp} \{ (\varepsilon_i-\mu_i)/ T \} +1},
\end{eqnarray}
where  $\varepsilon_i$ and $f_i(p)$ are the single-particle energy and the Fermi-Dirac distribution function, respectively, whereas subscript $i$ denotes the particle species, $i=n,p,\Lambda,\Sigma^-$.
We set each degeneracy factor $g=2$, and adopt each mass as $m_n=m_p=939$ MeV, $m_\Lambda=1115.7$ MeV, and $m_{\Sigma^-}=1197.4$ MeV.
Note that $\varepsilon_i$ includes the interaction energy $U_i$ as well as the kinetic energy \cite{baldo99, baldo04},
\begin{eqnarray}
\varepsilon_i = \sqrt{m_i^2 + p^2 } + U_i.
\end{eqnarray}

Finally, we get the free-energy density $\mathcal{F}$ as follows,
\begin{eqnarray}
\mathcal{F}_H = \sum_i  \left[ \cfrac{g}{(2\pi)^3} \int^\infty_0 \sqrt{m_i^2+p^2} f_i(p) ~4 \pi p^2 dp + \cfrac{1}{2} ~ U_i n_i \right]-Ts_H ,
\label{eq:free}
\end{eqnarray}
where $s_H$ is the entropy density calculated from
\begin{eqnarray}
s_H = - \sum_i \cfrac{g}{(2\pi)^3}
\int^\infty_0 \{ f_i(p) {\rm ln}f_i(p) +(1-f_i(p)) {\rm ln}(1-f_i(p)) \} ~ 4 \pi p^2 dp.
\end{eqnarray}

The total pressure for the uniform hadron phase is given by
\begin{eqnarray}
P_H= \sum_i \mu_i n_i   -  \mathcal{F}_H.
\end{eqnarray}

\subsection{Quark Phase; Bag model}

Next, we explain how to treat the quark phase based on the thermodynamic bag model.
In this model, we get the number density $n_Q$, the pressure $P_Q$, and the energy density $\epsilon_Q$,
\begin{eqnarray}
n_Q&=&\sum_q \cfrac{g}{(2\pi)^3} \int^\infty_0 f_q(p) 4 \pi p^2 dp, \\
\epsilon_Q&=&\sum_q \cfrac{g}{(2\pi)^3} \int^\infty_0 \sqrt{m_q^2 + p^2 } f_q(p) 4 \pi p^2 dp +B, \\
s_Q &=& - \sum_q \cfrac{g}{(2\pi)^3}
\int^\infty_0 \{ f_q(p) {\rm ln}f_q(p) +(1-f_q(p)) {\rm ln}(1-f_q(p)) \} ~ 4 \pi p^2 dp \\
\mathcal{F}_Q&=& \epsilon_Q -Ts_Q \\
P_Q&=& \sum_q \mu_q n_q   -  \mathcal{F}_Q.
\end{eqnarray}
for uniform quark matter,
where $f_q(p)$ is the Fermi-Dirac distribution function of the quark $q$ ($=u,d,s$), $m_q$ its current mass,
and $B$ the energy-density difference between
the perturbative vacuum and the true vacuum, i.e., the bag constant is set to be 100~MeV$\cdot$fm$^{-3}$. The baryon density $n_B$ is given by $(n_u+n_d+n_s)/3$.

Since there are still many uncertainties for the vacuum property, we here also adopt another type of $B$ by the density dependent bag model~\cite{nicotra06b}. In this model, we set the vacuum energy $B(n_B)$ as
\begin{eqnarray}
 B(n_B) = B_\infty +(B_0 - B_\infty) \exp \left[ -\beta \frac{n_B}{n_0} \right] \nonumber
 \label{eq:01}
 \end{eqnarray}
with $B_\infty = 50$ MeV$\cdot$fm$^{-3}$, $B_0=400$ MeV$\cdot$fm$^{-3}$, and $\beta=0.17$. Here, we set the saturation density as $n_0=0.17$ fm$^{-3}$. This  model has been proposed in Ref.\ \cite{nicotra06b}, on the basis of experimental results obtained at CERN SPS on formation of a quark-gluon plasma.\\

\begin{figure}[hbt]
\vskip 0.45in
\begin{center}
\includegraphics[width=0.55 \textwidth]{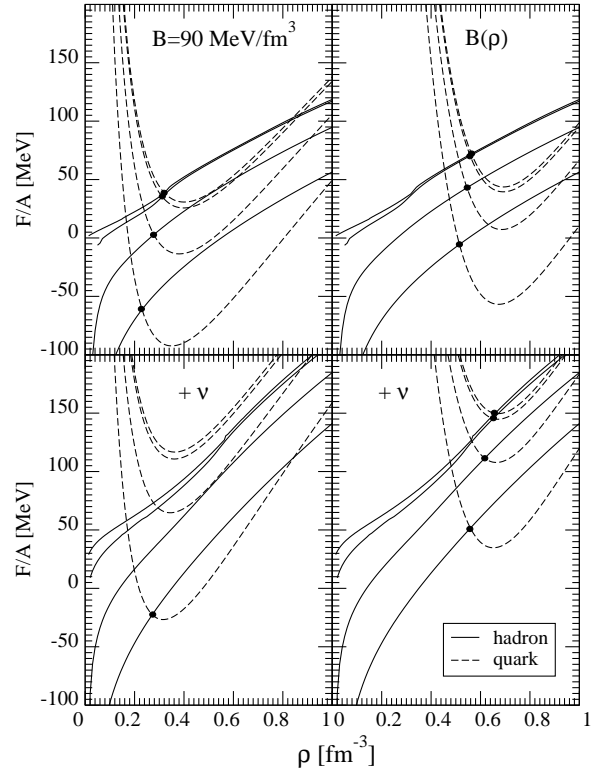}
\end{center}
\caption{Figure from Ref.\cite{nicotra06b}. Free energy per baryon as a function of baryon density of beta-stable hadron matter (solid lines) and quark matter (dashed lines) with a bag constant $B = 90$ MeV/fm$^3$ (left panels) or a density-dependent bag parameter (right panels) with (lower panels) and without (upper panels) neutrino trapping at different temperatures $T =$ 0, 10, 30, 50 MeV (upper to lower curves).}
\label{fig:nico}
\end{figure}

Here, we show free energy per baryon $F/A$ as a function of baryon density for hadron matter and quark matter at different temperature ($T=0, 10, 30, 50$ MeV) in Fig.~\ref{fig:nico}. We adopted the thermodynamic bag model ($B=90$ MeV/fm$^3$) as the quark matter for left panels, but the density dependent bag model for the right panels. The  lower panels show the case with neutrino trapping, $Y_L=0.4$, while upper panels for without. Note that the values of $F/A$ of hadrons are lower than the ones of quark matter at neutrino trapped case at $T=0, 10, 30$ for any density. Therefore it is not appropriate for our assumption that the HQ transition of the first order occurs at high density region. Hence we adopt the density-dependent bag model for the lepton trapping case in this article.

We note that the bag model should be quite simple to describe the quark matter.
We will update the quark model to more sophisticated models such as NJL model~\cite{burgio08, yasutake09a, masuda12}, pNJL model~\cite{fukushima08a, fukushima08b, kashiwa09, kouno10, lourenco11, lourenco12}, or Dyson-Schwinger model~\cite{chen09}, etc..

\subsection{The Maxwell Construction and the Gibbs Conditions}
\label{subsec:bulk}

\subsubsection{Hadron-quark Phase Transition as a Pure System Under the Maxwell Construction}

In this subsection, we introduce the mixed phase to discuss the phase equilibrium between hadron and quark phases.
We can expect the HQ transition to appear in various circumstance, supernovae, protoneutron stars, cold neutron stars or neutron-star mergers, where temperature may become high or neutrinos may be  trapped. So we consider here a general situation, where exist hadrons such as nucleons and hyperons and leptons such as electrons and neutrinos.

First, we consider the HQ phase transition without considering the finite-size effects which is called the bulk calculation. The naive treatment of the phase equilibrium in the first-order phase transition should be the application of the Maxwell construction.
In this treatment for neutron-star matter we implicitly assume that electric charge is always locally neutralized in each phase and lepton number is equal to each other; only baryon number is different between two phases. So the Maxwell construction means that the matter is regarded as a pure system, discarding the electromagnetic charge and lepton number.

Relaxing the local charge-neutrality condition and equality of lepton number, we must impose the Gibbs conditions in the proper treatment of the HQ transition in neutron stars. 
In this treatment, we impose the balance of the chemical potentials of quarks and baryons as well as the pressure balance and thermodynamical equilibrium between two phases.
The latter two conditions are obvious,
\begin{eqnarray}
P_H=P_Q , ~~~~~~~ T_H = T_Q ,
\end{eqnarray}
where $P_H$~($P_Q$), $T_H$~$(T_Q)$ are pressure and temperature of hadron~(quark) phase.
The former conditions are reduced to only one within the Maxwell construction for the baryon-number chemical potential $\mu_B$
\footnote{Another version of the Maxwell construction may be possible~\cite{pag}, where the effects of the lepton chemical potentials on $\mu_B$ are taken into account.}
,
\begin{eqnarray} \label{eq:maxwell}
\mu_{B}({\rm hadron~ phase}) = \mu_{B}({\rm quark~ phase}),
\end{eqnarray}
where the baryon chemical potential in the hadron~(quark) phase is given by $\mu_{B} = \mu_n $~($\mu_{B} =  \mu_u + 2\mu_d$).
Here, $\mu_n$, $\mu_u$, and $\mu_d$ is neutron, u-quark, and d-quark chemical potential, respectively.

We also assume the beta equilibrium among the particles as
\begin{eqnarray}
 && \mu_u = \cfrac{1}{3} \mu_B + \cfrac{2}{3} \mu_{C}^{Q},
~~~~~\mu_d = \mu_s = \cfrac{1}{3} \mu_B - \cfrac{1}{3}  \mu_{C}^{Q},
\label{eq:Qeq}
\end{eqnarray}
for the quark phase,
\begin{eqnarray}
 && \mu_n = \mu_\Lambda = \mu_B,
 ~~~~~ \mu_p =\mu_B + \mu_{C}^{H},
 ~~~~~ \mu_{\Sigma^-} + \mu_p = 2 \mu_B,
 \label{eq:Heq}
\end{eqnarray}
for the hadron phase, where the quark chemical potentials are shown as $\mu_{u, d , s}$ for u, d, s-quarks, the hadron chemical potentials $\mu_{n, p , \Lambda, \Sigma^-}$ for $n, p , \Lambda, \Sigma^-$. In these equations, we have also introduced the two kinds of variables to understand the mechanism of the HQ phase transition systematically; the lepton chemical potential $\mu_{L}^{ H(Q)}$ and the charge chemical potential $\mu_{C}^{H(Q)}$ are given as
\begin{eqnarray}
\mu_{L}^{H(Q)} = \mu_{\nu_e}^{H(Q)},
~~~~~~~~~~
\mu_{C}^{H(Q) }= \mu_L^{H(Q)} - \mu_{e}^{H(Q)},
\end{eqnarray}
in hadron~(quark) matter. Here the lepton chemical potentials are shown as $\mu_{e}^{ H(Q)}$ and $\mu_{\nu_{e}}^{H(Q)}$ for electrons and electron type neutrinos, respectively.
Note that $\mu_L$ and $\mu_{C}$ are not equal to each other between two phases; $\mu_{L}^{ H} \neq \mu_{L}^{Q}$ and $\mu_{C}^{ H} \neq \mu_{C}^{ Q}$.  Under this construction, we assume only the equality of baryon chemical potentials $\mu_B$ between two phases shown as Eq.(\ref{eq:maxwell}). Hence the matter is regarded as a pure system in this treatment. As a result, the EOS has a density jump in general.

\subsubsection{Hadron-quark Phase Transition as a Ternary System Under the Gibbs Conditions}

So far the charge equilibrium is completely discarded between two phases: the above assumption uses the charge neutrality condition in the local form, while it must be attained as a whole system.  Consequently, the Maxwell construction arises as a consequence of further strong constraint. In such treatment, the lepton number is also assumed to be conserved locally. In the proper treatment of the HQ transition we must consider both phases as ternary systems specified by the three conserved quantities, electromagnetic charge, lepton number, as well as baryon number.

 In other words, we must add two chemical equilibrium-conditions into  Eq.(\ref{eq:maxwell}) as
\begin{eqnarray} \label{eq:GCN}
\mu_{B}^H = \mu_{B}^Q,~~~~~~~~~
\mu_{L}^H = \mu_{L}^Q,~~~~~~~~~
\mu_{C}^H = \mu_{C}^Q.
\end{eqnarray}
Here, there are three identical components;  the baryon number (which corresponds to $\mu_B$), charge ($\mu_C$), and lepton number ($\mu_L$).
The Maxwell construction is then meaningless and we must use the Gibbs conditions to describe the phase equilibrium in a thermodynamically consistent way.
Hence, the EOS dose not have density jump although we assume that the phase transition is of the first order.

Since we impose the global charge neutrality~(GCN), the total charge as well as baryon-number are shared by the two phases, conservation should be satisfied as following;
\begin{eqnarray}
n_B &=& \chi n_{B}^{Q}+(1-\chi) n_{B}^{H}~~~,\\
Y_C ~ n_B &=& \chi ~ Y_{C}^{Q} ~ n_{B}^{Q}+(1-\chi)~ Y_{C}^{H}~ n_{B}^{H}~~~,\label{eq:charge}\\
Y_L ~ n_B &=& \chi ~ Y_{L}^{Q} ~ n_{B}^{Q}+(1-\chi)~ Y_{L}^{H}~ n_{B}^{H}~~~,
\end{eqnarray}
where $n_B, n_{B}^{Q}, n_{B}^{H}$, and $\chi$ are respectively the total baryon number density, the  baryon number density in hadron matter, the effective baryon number density in quark matter, and the volume fraction which indicates the volume of the quark phase divided by the total volume.  Here, $n_{B}^{Q}$ is one third of the total quark number density. The values of $Y_C$  and $Y_L$ are the total fraction of positive particles and leptons, whereas $Y_{C(L)}^{Q}$ and $ Y_{C(L)}^{H}$ denote the fractions of positive particles~(leptons) per baryon in quark matter, and in hadron matter.
Note that $Y_{C(L)}^{Q}= Y_{C(L)}^{H}$ are assumed in the previous treatment (\ref{eq:maxwell}) instead of $\mu_{C(L)}^{Q}=\mu_{C(L)}^{H}$.

\bigskip

\subsubsection{Hadron-quark Phase Transition as a Binary System with the Local Charge Neutrality}

Recently the hybrid types of the Maxwell construction and the Gibbs conditions have been suggested,  depending on the physical situation~\cite{pag}. One of them is the Gibbs conditions assuming the local charge neutrality~(LCN), $Y_{C}^{Q}= Y_{C}^{H}=0$.
In this treatment, the chemical equilibriums are described as follow;
\begin{eqnarray} \label{eq:LCN}
\mu_{B}^H = \mu_{B}^Q,~~~~~~~~~
\mu_{L}^H = \mu_{L}^ Q.
\end{eqnarray}
We shall see LCN is turned out to be achieved in some cases as in the protoneutron stars~(subsection \ref{subsec:neutrino}) .

\bigskip
\bigskip

\begin{figure}[hbt]
\begin{center}
\includegraphics[width=0.49\textwidth]{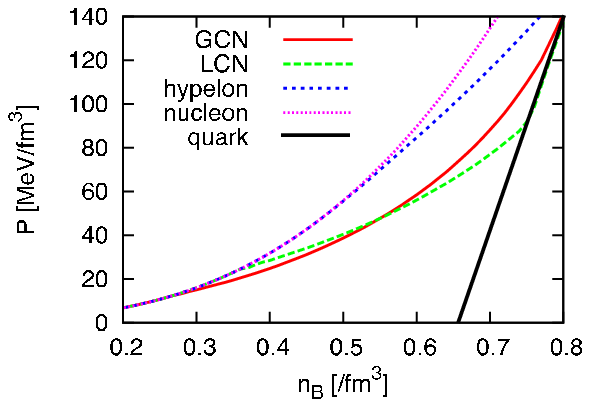}
\includegraphics[width=0.49\textwidth]{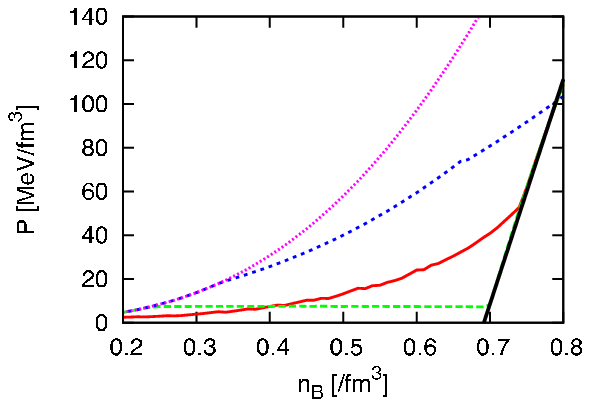}
\end{center}
\caption{Equation of sate considering the mixed phases imposed the Gibbs conditions assuming GCN and LCN, which is labeled or `GCN' and `LCN'. For comparison, we also shows the uniform matter of quark matter~(labeled as `quark') and hadron matter. Here, the EOS labeled as `hyperon' shows the hadron matter including hyperons, while the EOS labeled as `nucleon' is without hyperons. Left panel shows the EOS at isentropic, $s=2 [/k_B]$, and at $Y_L= 0.35$. Right panel is the same with left one, but for at zero temperature without neutrinos.}
\label{fig:bulkeos}
\end{figure}

In the last of this subsection, we show some examples of EOS adopted the Gibbs conditions for the HQ mixed phase in Fig.\ref{fig:bulkeos}, where we compare both conditions of GCN and LCN. The left panel is at finite isentropic case $s=2~[/k_B]$, and $Y_L=0.35$ assuming the matter in protoneutron stars.
One can see that the proper treatment of the Gibbs conditions (GCN) gives almost the same result as that in the hybrid treatment of the Gibbs conditions with a further constraint, LCN, in this case.
The right panel is at zero temperature without neutrinos assuming the case in the neutron stars. Note that, without neutrinos, the Gibbs conditions with LCN give completely the same result with the Maxwell construction. Consequently, the EOS has a density jump as shown in the right panel in Fig.\ref{fig:bulkeos} while pressure smoothly increases in GCN. We also show the uniform EOSs of quark matter and hadron matter in these figures.

\subsection{Finite-size Effects and ``Pasta'' Structures}

\begin{figure}[hbt]
\begin{center}
\includegraphics[width=0.9\textwidth]{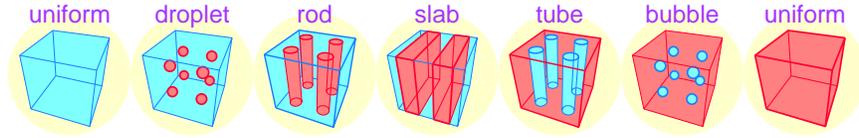}
\end{center}
\caption{Schematic picture of ``pasta'' structures.
When red phase appears in the blue phase, the red phase
form droplets in the sea of blue.
With increase of the volume fraction of red phase in the blue phase,
the shape of the red phase changes from droplet to rod,
slab, tube, and to bubble before pure red phase appears.}
\label{figSchematicPasta}
\end{figure}

We have presented general features of the non-congruent transition in a clear fashion with a help of the bulk calculation, discarding the Coulomb interaction and surface effects
called the finite-size effects. However, the bulk calculation is insufficient to figure out the essential aspects of the non-congruent transition.
Actually we can see the emergence of imhomogeneous structure with various geometrical shapes, called pasta (see Fig.~\ref{figSchematicPasta}),
by the cooperative effect of the Coulomb interaction and the surface energy, which is similar to the nuclear pasta at subnuclear densities \cite{Lorenz1993, Oyamatsu1993, SOT1995, maru2005, new, oka}.
The charge screening effect, which is a typical many-body effect in the presence of the Coulomb interaction, plays an important role in this context to
make the pasta structures unstable, which is called mechanical instability~\cite{voskre}.
The evaluation of the free energy of the pasta structures can be carried out by using the Wigner-Seitz approximation;
dividing the whole space into equivalent unit cells, we solve the coupled equations for density profiles of particles and the Poisson equation
in a selfconsistent way within the unit cell \cite{WScell}. We further use the Thomas-Fermi (local density) approximation for each particle.

Assuming various geometrical structures, we must numerically search the minimum of the free energy
\begin{eqnarray}
F =E_V  + F_e + F_{\nu_e}+ E_C + E_S
\label{eq:03}
\end{eqnarray}
changing the volume ratio of quark matter to hadron matter, and the size of Wigner-Seitz cell $R_W$.
Here the bulk energy $E_V$ gives a main contribution,
\begin{equation}
E_V=\int_{V_H} dr^3 \mathcal{F}_H[n_i]
 + \int_{V_Q} dr^3 \mathcal{F}_Q[n_a],
\end{equation}
where $\mathcal{F}_H$ and $\mathcal{F}_Q$ comes from the kinetic energy and strong interaction.
The free energies of electrons, $F_e$, and neutrinos, $F_{\nu_e}$, are simply calculated by the use of the Fermi-Dirac distribution functions.
$E_C$ denotes the interaction energy due to the Coulomb force.
 The treatment of the Coulomb potential may deserve some discussions,
since it includes a delicate problem and sometimes has given rise to a confusion \cite{voskre}.
First of all we must notice that charge chemical potential $\mu_C$ itself is not physical quantity
since it is gauge dependent in the presence of the Coulomb interaction $V_{\rm Coul}$.
Only their combination of $\mu_C-V_{\rm Coul}$ is gauge invariant.
Consider, for example, electrons without any neutrino. Then the electron chemical potential $\mu_e$
is equal to $\mu_C$, so that the Gibbs condition means $\mu_e$ equals to each other in the mixed phase.
On the other hand, the electron number density $n_e$ should be represented as $(\mu_e-V_{\rm Coul})^3/(3\pi^2)$
in the gauge invariant way. Thus if $V_{\rm Coul}$  spatially varies in the mixed phase, each phase can have different
number of electrons even if $\mu_e$ is the same in the both phases.
In the extreme case $V_{\rm Coul}$ may resemble the step function
and we have a simple picture of uniform electron density in both phases with different numbers.
Note that there arises no such problem in the absence of the Coulomb potential as in the bulk calculation:
the difference of the electron number directly reflects in the difference of the electron chemical potential in this case.

Here, we do not consider all kinds of anti-particles for simplicity.
After the minimization of the free energy, we can get the thermodynamic quantities, such as pressure, entropy, etc..
The notation $E_S$ stands for the surface energy which comes from
a sharp interface between the hadron and quark phases with a fixed surface tension $\sigma$.
Although the surface tension of the HQ interface is poorly known,
some theoretical estimates based on the MIT bag model
for strangelets \cite{jaf} and lattice gauge simulations at finite temperature \cite{latt} have suggested a range of $\sigma \approx 10$--$100\;\rm MeV\cdot fm^{-2}$.
We will discuss the effects of the surface tension on the pasta structures in subsection \ref{subsec:surf}.

\begin{figure}[hbt]
\begin{center}
\includegraphics[width=0.55\textwidth]{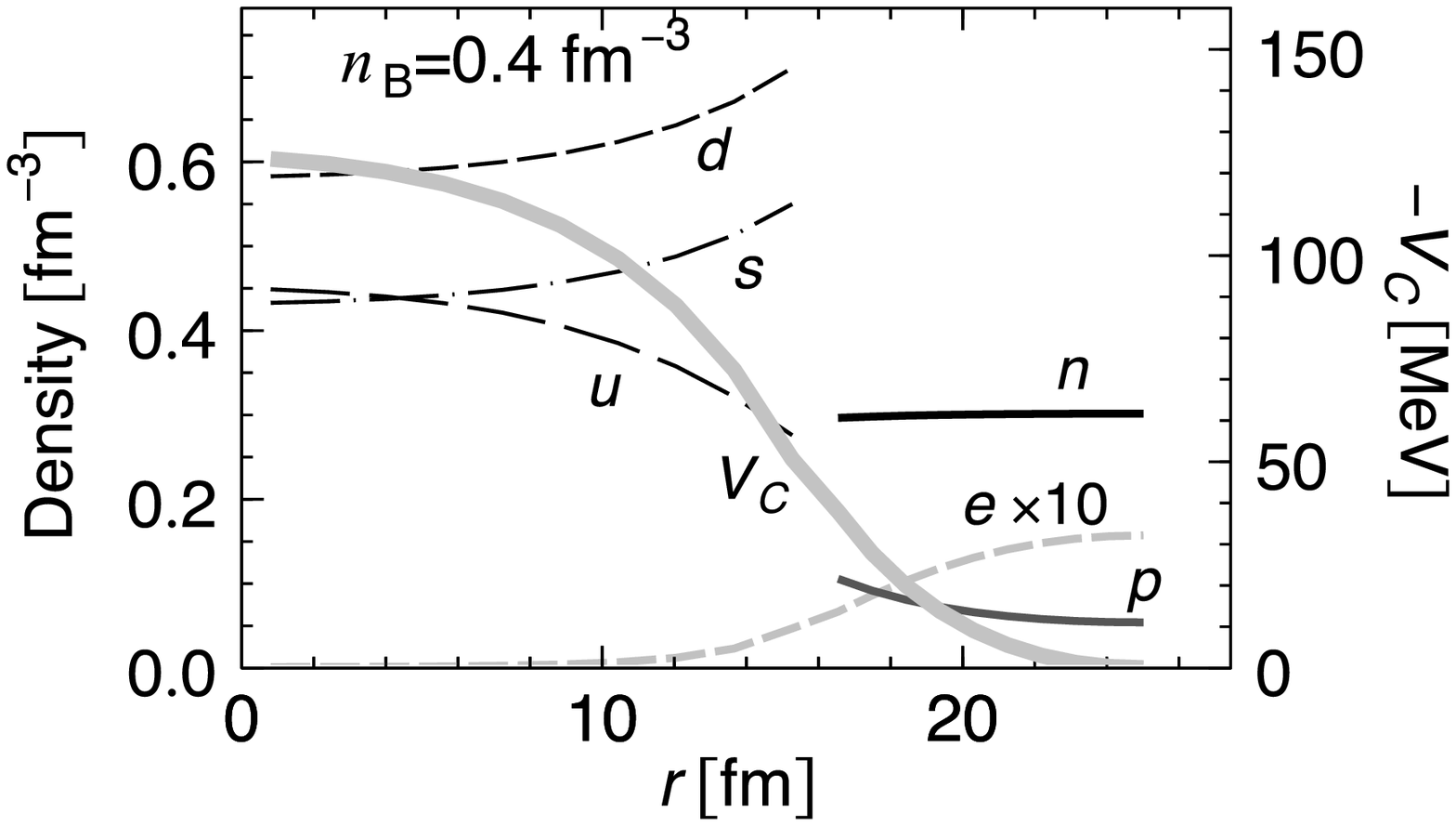}
\end{center}
\caption{
Density profiles 
and the Coulomb potential $V_C$ 
within a 3D (quark droplet) Wigner-Seitz cell of the mixed phase at $n_B=0.4$ fm$^{-3}$. The cell radius and the droplet radius are $R_W=26.7$ fm and $R=17.3$ fm, respectively.}
\label{figProf}
\end{figure}

Fig.~\ref{figProf} shows an example of the resulting density profiles in the droplet phase for $n_B=0.4$ fm$^{-3}$ at zero temperature~\cite{hyp07}.
One can see the non-uniform density distributions of particle species
together with the non-vanishing Coulomb potential.
The quark phase is negatively charged, so that
$d$ and $s$ quarks are repelled to the phase boundary,
while $u$ quarks gather at the center.
The protons in the hadron phase are attracted by the negatively charged
quark phase, while the electrons are repelled.

\subsection{Hyperon Suppression in the Mixed Phase}

\begin{figure}[hbt]
\begin{center}
\includegraphics[width=0.9\textwidth]{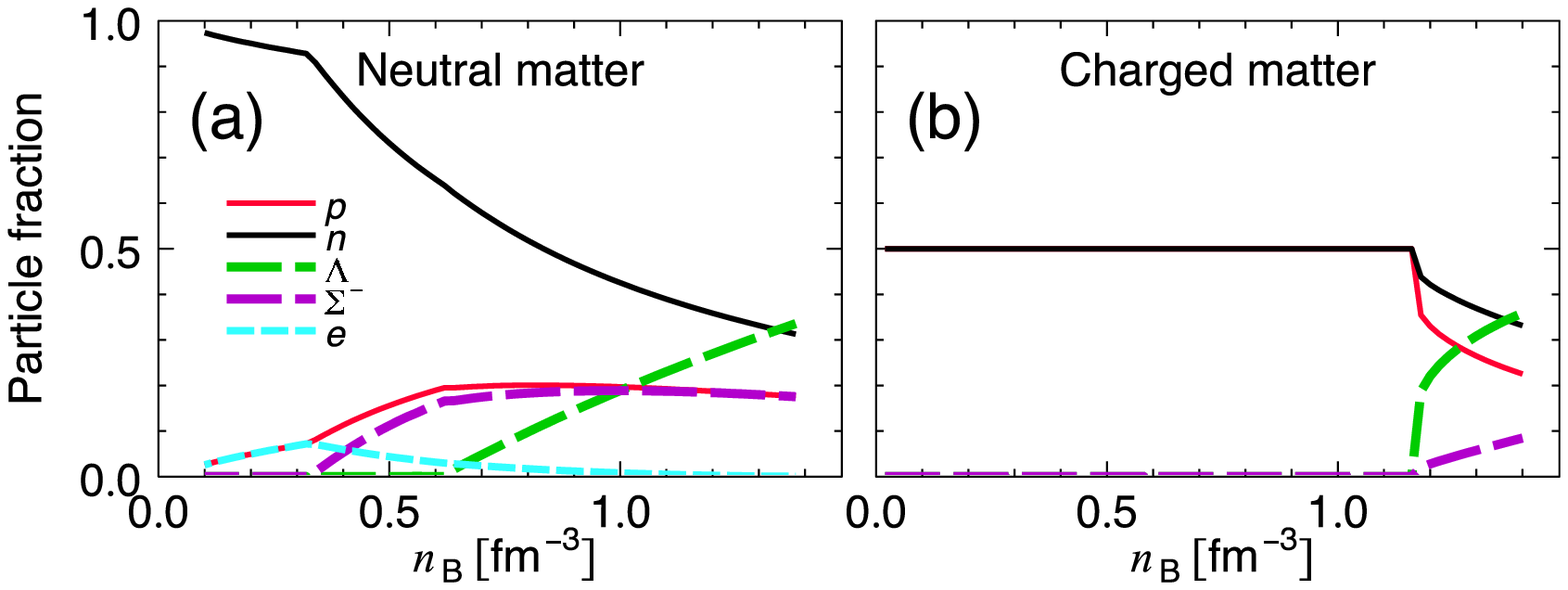}
\end{center}
\vspace{-3mm}
\caption{
(a) Particle fractions of neutral matter with electrons. 
(b) The same quantity for charged matter without electrons,
the low-density part of which corresponds to symmetric nuclear matter.
}
\label{figRatioUnif}
\end{figure}

Though we allow the freedom of hyperon mixture, hyperons do not actually
come into the mixed phase. The suppression of hyperon mixture in the mixed phase
is due to the fact that the hadron phase is positively charged.
As shown in Fig.\ \ref{figRatioUnif},
hyperons ($\Sigma^-$) appear in charge-neutral hadronic matter
at low density 
to reduce the Fermi energies of electron and neutron.
Without the charge-neutrality condition, on the other hand,
there is symmetric nuclear matter at lower density
and hyperons will be mixed there 
at higher density due to the large hyperon masses.
Generally speaking hyperons are hardly mixed in the positively charged matter.
Thus, the mixture of hyperons is suppressed in the mixed phase where the hadron phase
is positively charged.

\subsection{Effect of the Surface Tension}
\label{subsec:surf}

\begin{figure}[hbt]
\begin{center}
\includegraphics[width=0.48\textwidth]{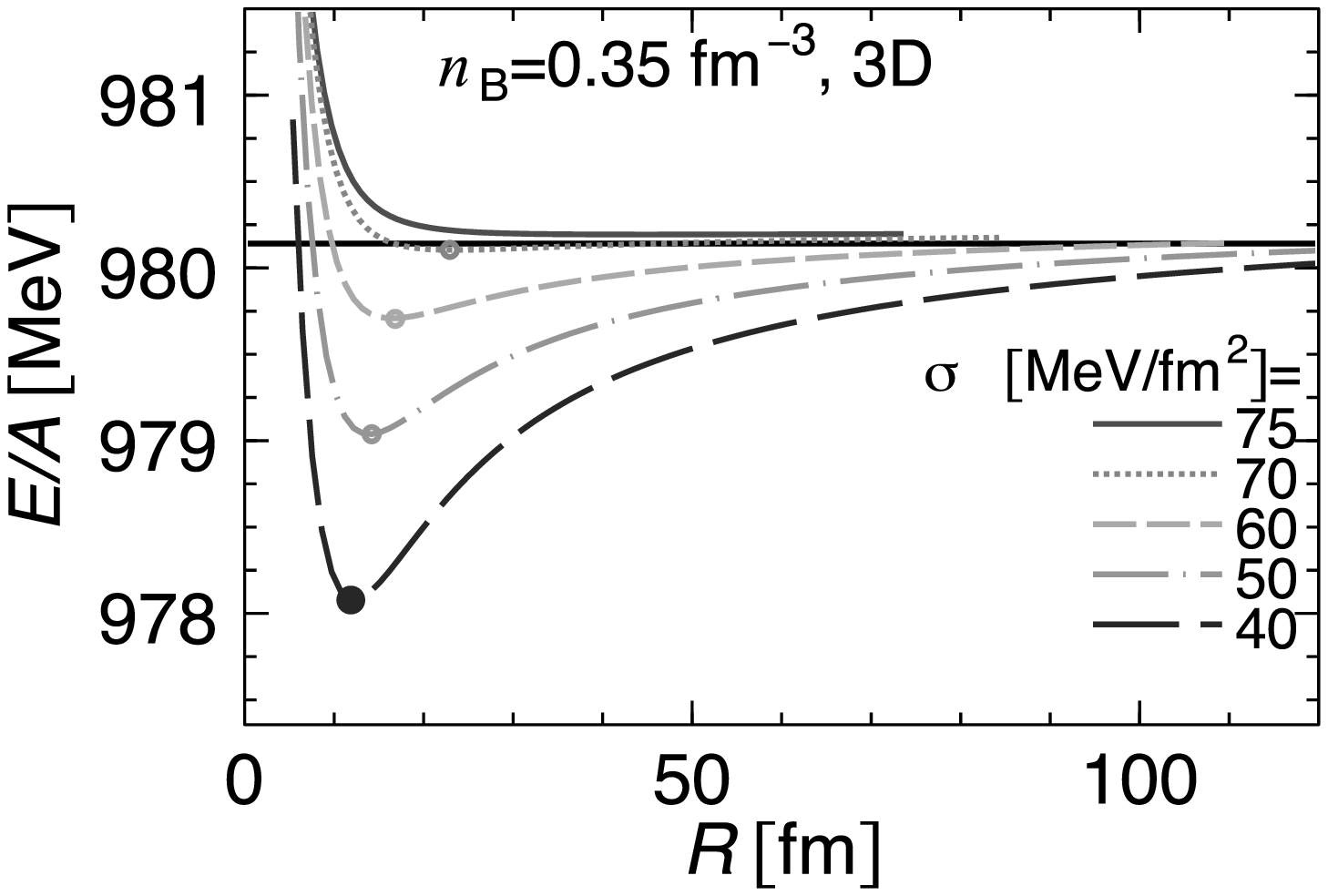}
\end{center}
\caption{
Droplet radius ($R$) dependence of the energy per baryon
for fixed baryon density $n_B=0.35$ fm$^{-3}$
and different surface tensions. The temperature $T$ is zero for all cases.
The quark volume fraction $(R/R_W)^3$ is fixed for each curve.
Dots on the curves show the local energy minima.
The black line shows the energy of the Maxwell construction case.
}
\label{figRdep}
\end{figure}

Let us consider the role of the surface tension on the mixed phase.
As already mentioned, the surface tension in the interface of the hadron and quark phases is poorly known.
If one uses a smaller surface tension parameter $\sigma$,
the energy gets lower and the density range of the mixed phase gets wider.
The limit of $\sigma=0$ MeV$\cdot$ fm$^{-2}$ leads to a bulk application of
the Gibbs conditions without the Coulomb and surface effects.
On the other hand, using a larger value of $\sigma$,
we find that the scale of the inhomogeneity becomes large
and the EOS gets closer to that of the Maxwell construction case.
Beyond a limiting value of $\sigma \approx 65\;\rm MeV\cdot fm^{-2}$
the structure of the mixed phase becomes mechanically unstable~\cite{voskre}:
for a fixed  volume fraction $(R/R_W)^3$
the optimal values of $R$ and $R_W$ diverge
and local charge neutrality is recovered in the mixed phase,
where the energy density equals to that of the Maxwell construction
(see Fig.\ \ref{figRdep}).

Note that these roles of surface tension on the EOS is remarkable for the matter without neutrinos, i.e. a binary system,
where the pasta structure is rather stable.
If we consider the neutrino contamination, the matter should be treated as a ternary system. As we will discuss at subsection \ref{subsec:neutrino},
the EOS could not be close to the one of the Maxwell construction,
but to the one with the Gibbs conditions~(LCN) depending on the neutrino fraction.

\subsection{Thermal Effects}
\label{subsec:neutrino}
\begin{figure}[hbt]
\includegraphics[width=.5\textwidth]{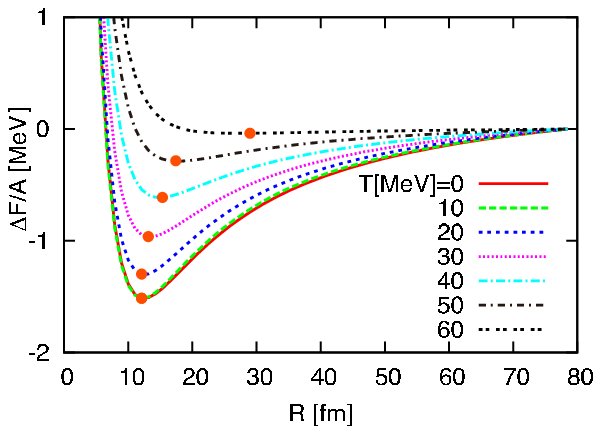}
\includegraphics[width=.5\textwidth]{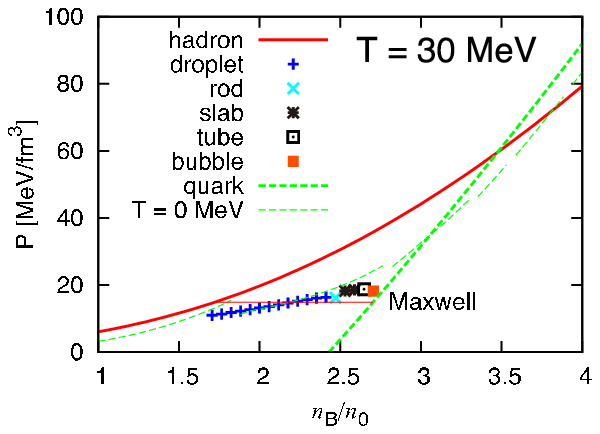}
\caption{Left panel shows that the radius $R$ dependence of the free-energy per baryon for the droplet case at $n_B = 2~n_0$ and different temperatures. The free energy is normalized by the value at $R \rightarrow \infty$. The filled circles on each curve shows the energy minimum. The results are for $B = 100$ MeV$\cdot$fm$^{-3}$, $\sigma= 40$ MeV$\cdot$fm$^{-2}$.
The right panel shows the EOS of the mixed phase (thick dots) in comparison with pure hadron and quark phases (thin curves) at $T = 30$ MeV. We also show, for comparison, the mixed phase by the Maxwell construction by thin solid line.}
\label{fig:ny02}
\end{figure}

In following we discuss the thermal and neutrino trapping effects on the HQ phase transition. In the left panel of Fig.\ \ref{fig:ny02} we show the dependence of the isothermal free-energy  per baryon on the cell size in Eq.~(\ref{eq:03}) for several temperatures at  $n_B = 2~n_0$, using the constant bag constant and the surface tension as $B=100$ MeV$\cdot$fm$^{-3}$ and $\sigma= 40$ MeV$\cdot$fm$^{-2}$, respectively~\cite{yas09}.
Note that the free energy per baryon $\Delta F/A$ is normalized value by subtracting the one at infinite $R_W$, for the comparison of thermal effects.
Here we assume the geometrical structure is the droplet and do not take into account neutrinos to see the genuine thermal effects.
The quark volume fraction $(R/R_W)^3$ is fixed to be the optimal value at $T = 0$ MeV for each curve.
From the this panel, we can see that the minimum point is shifted to a larger value of $R$ as temperature is increased, and eventually disappears for $T > 60$ MeV. The temperature dependence of the free energy comes from the Coulomb energy, the surface energy and the correlation energy.  The correlation energy is given by the difference in the bulk energy $E_V$
in Eq.~(\ref{eq:03}) \cite{voskre}, defined by
\beq
E_{\rm corr}\equiv E_V(n_i({\bf r}))-E_V(n_i^{I,II}).
\eeq
The rearrangement of the particle densities also induces a change in the volume energy $E_V$, which is composed of kinetic energy and strong-interaction energy. In the above calculation $E_V$ itself is intrinsically large but independent on $R$, since we have assumed uniform densities $n_i^{I,II}$ in each phase. Once the Coulomb interaction is properly taken into account, particle densities are no more uniform to produce a $R$ dependence in $E_V$.
The detailed analysis showed that $E_{\rm corr}$ gives negative values and behaves $R^{-1}$ for large $R$.
By comparing the temperature dependence of these contributions, we can see that the correlation energy is primarily responsible to the behavior of the minimum point. It means that the mixed phase becomes less stable as  temperature is increased ~\cite{yas09}.

In the right panel of Fig. \ref{fig:ny02}, we show the EOS (pressure vs baryon density) for the neutrino-free case at $T=30$ MeV. Clearly, the pressure becomes to close to the one given by the Maxwell construction at finite temperature, since the mixed phase becomes unstable and its density regime is thereby largely restricted. The mechanical instability of the pasta structure also arises in other geometrical shapes, which means that both $R$ and $R_W$ go to infinity with their ratio fixed. Thus we recover the picture of the phase equilibrium of two bulk matters, where the surface energy is irrelevant and the Coulomb energy vanishes due to the achievement of the local charge-neutrality. These features indicate that the EOS resembles the one given by the Maxwell construction.

\subsection{Neutrino Trapping Effects}
\label{subsec:neutrino}

\begin{figure}[hbt]
\centerline{
\includegraphics[width=.5\textwidth]{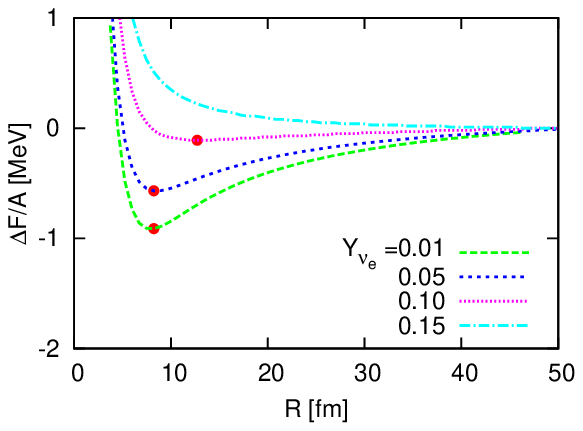}
\includegraphics[width=.46\textwidth]{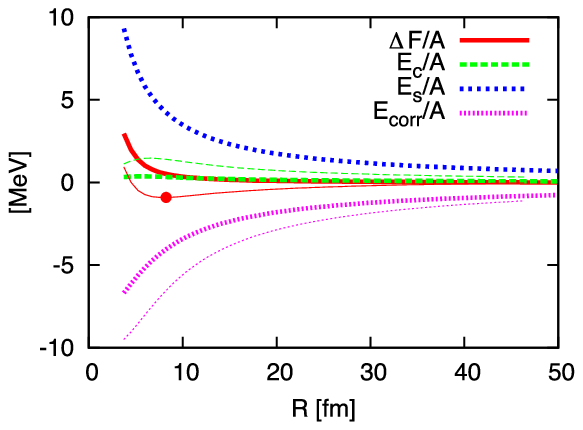}
}
\caption{\label{fig:ny03} Left panel is the same as the left panel of Fig.\ \ref{fig:ny02}, but for different neutrino fractions $Y_{\nu_e}$ at a given temperature, $T=10$ MeV.  In this case, we set baryon density as $n_B = 2.5~n_0$. The thick lines of the right panel are the case of $Y_{\nu_e} = 0.15$, the thin lines $Y_{\nu_e} = 0.01$.
The right panel shows each component of the free energy ($\Delta F/A$); e.g. the free energy, the Coulomb energy~($E_{\rm C}$/A), the surface energy~($E_{\rm S}$/A), and the correlation energy~($E_{\rm corr}$/A). Here they are shown per baryon.  }
\end{figure}

In Fig.\ \ref{fig:ny03}, we show the dependence of the free energy on neutrino trapping  in the mixed phase~\cite{yas12}. Here, we adopt the density-dependent bag, and fixed the quark volume fraction $(R/R_W)^3$ at $Y_{\nu_e} = 0.01$ for each curve. For $Y_{\nu_e} > 0.1$, the minimum point disappears. The right panel shows the each contribution to the free energy.  From this figure, we can see that both the correlation energy and the Coulomb energy mainly contribute to the behavior of the minimum point in the presence of neutrinos, while it is a only  the correlation energy for the temperature dependence.

\begin{figure}[hbt]
\centerline{
\includegraphics[width=.5\textwidth]{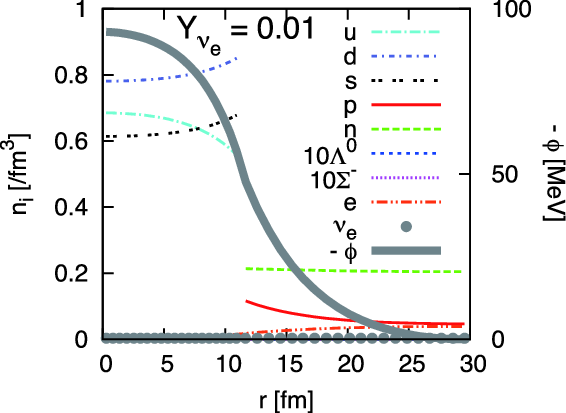}
\includegraphics[width=.5\textwidth]{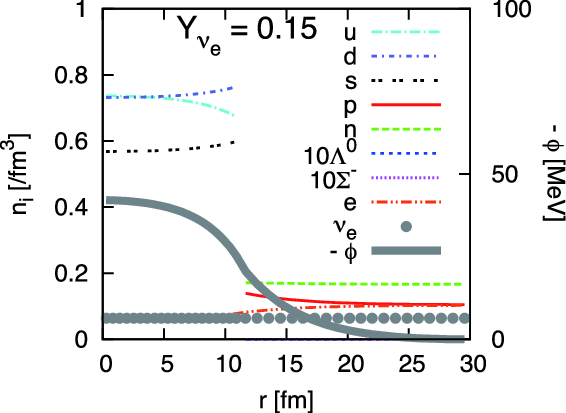}
}
\caption{\label{fig:ny04}Density profiles and the Coulomb potential $\phi$ within 1D (slab) for $n_B = 2.5~n_0$ at $T=10$ MeV. Here, the neutrino fractions are set to  $Y_{\nu_e}=0.01$~(left panel) and $Y_{\nu_e}=0.15$~(right panel). The cell sizes are $R_W = 30$ fm in these figures. The slab size are $R = 10.9 $ fm. $(R/R_W)^3$ is fixed to be the optimal value at $Y_{\nu_e} = 0.01$.}
\end{figure}

To elucidate the neutrino-trapping effect in the mixed phase, we show the density profiles and the Coulomb potential for the slab case in Fig. \ref{fig:ny04}.
For comparison, we use the same cell-size, and fix $R_W$ here.  High neutrino fraction enhances the number of electrons to satisfy the chemical equilibrium.
As a result, the Coulomb potential drastically changes in the neutrino trapping case as shown in this figure.

\begin{figure}[hbt]
\begin{center}
\includegraphics[width=15pc]{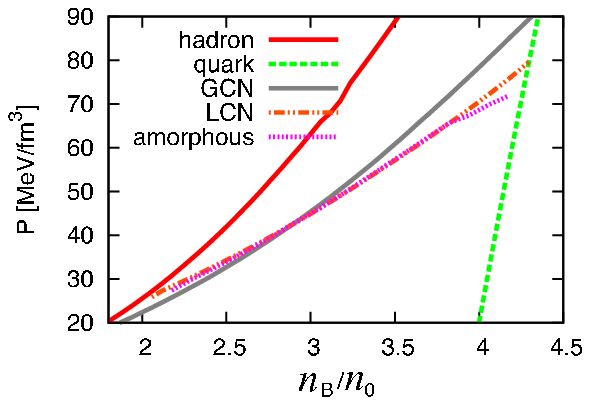}
\includegraphics[width=15pc]{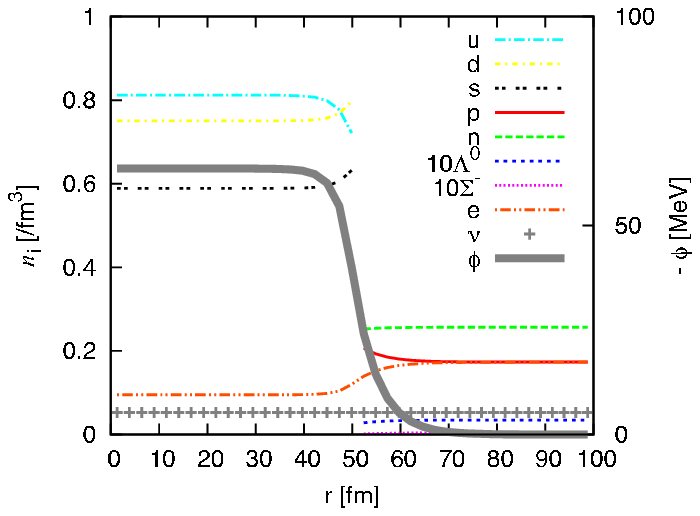}
\caption{Left panel shows the pressure for PNS matter of $T=30$MeV and $Y_L=0.4$. The nuclear density is denoted by $n_0$. Right panel shows the density profile and the Coulomb potential $3 n_0$ assuming the rod structure for the same PNS matter.
}
\vspace{-5mm}
\label{fig:P_LCN}
\end{center}
\end{figure}

In the left panel of Fig.~\ref{fig:P_LCN}, we show the  relation of the pressure and density for PNS matter~($T=30$MeV and $Y_L=0.4$) as a result of above calculations. For comparison, we also check the phase transition under the Gibbs conditions with GCN and LCN. As shown in the left panel of Fig.~\ref{fig:ny03}, there is no minimum point of the free energy for PNS matter, which means the crystalline structures of pasta are broken there to form gan amorphous stateh composed of two species of matter; the amorphous state should take a bicontinuous structure with a complex pattern, depending on the kinetics of its formation. In such case, the surface tension and the Coulomb interaction do not work to make any pasta structure (see the right panel of Fig.~\ref{fig:ny03}), so that the mixed phase may resemble the amorphous phase of quarks and hadrons.
For numerical reason, we can not calculate $R_W = \infty$.
Instead, we set $R_W = 100$ fm for amorphous state.

From the left panel of Fig.~\ref{fig:P_LCN} and the above discussion, the amorphous state of quarks and hadrons gives almost the same result with the bulk Gibbs calculation  with LCN. To understand this reason, we also show the density profile and the Coulomb potential $3 n_0$ assuming the rod structure for the same PNS matter in the right panel of Fig.~\ref{fig:P_LCN}. From this figure, we can see that the shape of the Coulomb potential looks like a step function. This is because the size of $R_W$ is very large ($R_W = 100$ fm); the surface effect becomes relatively small and the density profiles become almost uniform in each phase. Note again that the electron density is different in each case even if we impose the condition $\mu_e^{H}=\mu_e^{Q}$ in the presence of the Coulomb potential.

\bigskip
\bigskip

In the last of this subsection we summarize the above discussions for the HQ phase transition in Table 1.
Basically the matter should be treated as a ternary system specified by the three kinds of the conserved quantities, baryon number, electric charge and lepton number. The proper treatment of the phase transition and the mixed phase proceeds by applying the Gibbs conditions. There are some kinds of variation instead of the proper treatment: the bulk Gibbs calculation (GCN) may be performed by neglecting the finite-size effects. The bulk Gibbs calculation has been also done with the use of the assumption of local charge neutrality (LCN). One may regard the matter as a binary system since the chemical equilibrium is required only for
baryon and lepton numbers in this treatment. In the realistic treatment of the mixed phase, the finite-size effects are indispensable. When the finite-size effects are taken into account, we see the emergence of the inhomogeneous structure called pasta.
We have seen that the proper treatment based on the Gibbs conditions gives almost the same EOS as the one given by the bulk Gibbs calculation with LCN in PNS matter. This is because the  scale of inhomogeneity becomes large, e.g. $R_W\rightarrow \infty$, and an amorphous state is formed, where the finite-size effects are irrelevant and local charge neutrality is also attained. This is due to the mechanical instability by the neutrino trapping effect.
Thus we can say the ternary system is effectively reduced to a``binary" system. We attach ``binary" in the table to stress this situation.

The situation somewhat changes for various cases. Neutron star (NS) matter
is a cold crystallized matter and includes no neutrino. So it should be treated as a binary system specified by the baryon number and electric charge. We have seen that it is effectively reduced to a ``pure" system due to the mechanical instability for sufficiently large surface tension or high temperature; the scale of the inhomogeneity becomes large again and local charge neutrality is approximately achieved. Consequently the EOS resembles the one given by the Maxwell construction. The NS-NS mergers corresponds to the latter case, and we named ``merger (MG)" matter. In this sense we listed the system as ``pure" in the table
 for these cases. Note that there is one important difference between these cases and PNS matter, while the situation looks similar to each other; there is no amorphous state in these cases and appears a density jump inside compact stars instead.

\begin{table}[hbt]
\begin{center}
\caption{\label{tab:00}
Comparison of conditions for the HQ phase transition.}
{\scriptsize
\begin{tabular}{lccccc}
\hline
          & finite-   & globally      & locally          & \\
         &  size       & conserved & conserved  &  equilibrium conditions & system\\
         & effects  & variables    & variables       & \\
\hline
\hline
Maxwell        & No & $n_B$ & $Y_L, Y_C$              & $\mu_{B}^{ H}=\mu_{B}^{Q}$ & pure \\
\hline
bulk Gibbs\\
(GCN)           & No & $n_B, Y_L, Y_C$ & & $\mu_{B}^{H}=\mu_{B}^{Q},
~\mu_{L}^{H}=\mu_{L}^{Q},~ \mu_{C}^{ H}=\mu_{C}^{Q}$ & ternary\\
 \hline
bulk Gibbs\\
(LCN)            & No & $n_B, Y_L$  & $Y_C$ & $\mu_{B}^{H}=\mu_{B}^{Q},
~\mu_{L}^{H}=\mu_{L}^{Q}$ & binary \\
\hline
pasta            & Yes  & $n_B, Y_L, Y_C$  & &$\mu_{B}^{ H}=\mu_{B}^{Q},
~\mu_{L}^{H}=\mu_{L}^{Q},~\mu_{C}^{ H}=\mu_{C}^{Q}$ & ternary \\
\hline
NS matter \\
pasta   & Yes & $n_B, Y_C$  & &$\mu_{B}^{ H}=\mu_{B}^{Q},
~ \mu_{C}^{ H}=\mu_{C}^{Q}$ & binary \\
 \hline
NS matter\\
pasta~(large $\sigma$) & Yes  & $n_B, Y_C$  & &$\mu_{B}^{ H}=\mu_{B}^{Q},
~\mu_{C}^{ H}=\mu_{C}^{Q}$ & ``pure" \\
\hline
MG matter  \\
pasta   & Yes & $n_B, Y_C$  & &$\mu_{B}^{ H}=\mu_{B}^{Q},
~ \mu_{C}^{ H}=\mu_{C}^{Q}$ & ``pure" \\
\hline
PNS matter\\
amorphous & Yes & $n_B, Y_L, Y_C$  & & $\mu_{B}^{H}=\mu_{B}^{Q},
~\mu_{L}^{H}=\mu_{L}^{Q},~ \mu_{C}^{H}=\mu_{C}^{Q}$ & ``binary" \\
\hline
\end{tabular}
}
\end{center}
\end{table}


\section{Possibility to Probe the Properties of the Mixed Phase}
\label{astro}

As mentioned above, possible existence of the HQ mixed phase in high density region can be theoretically expected.
However, it might be quite difficult on the Earth to reveal the properties of such an extreme state. The one of the possibilities to overcome this difficulty may be the direct observation of neutron stars. In fact, since the stellar structure of neutron stars strongly depends on their interior properties, one could obtain the information of the stellar interior via the observations of stellar mass $M$, radius $R$, and compactness defined as $M/R$.

Once the EOS of matter is obtained, we can extract the density profile
(density as a function of radial distance from the center) and consequently
the mass and the radius of compact stars.
This is done by numerically solving the Tolman-Oppenheimer-Volkoff (TOV)
equation \cite{sha},
\begin{eqnarray}
  {dp\over dr} &=& -{ G m \epsilon \over r^2 } \,
  {  \left( 1 + {p / \epsilon} \right)
  \left( 1 + {4\pi r^3 p / m} \right)
  \over
  1 - {2G m/ r} } \:,\qquad
\label{tov1}\\
  {dm \over dr} &=& 4 \pi r^2 \epsilon \:,
\label{tov2}
\end{eqnarray}
being $G$ the gravitational constant.
Starting with a central density $\epsilon(r=0) \equiv \epsilon_c$,
one integrates out Eqs.~(\ref{tov1}) and (\ref{tov2}) until
the surface pressure becomes to zero.
This gives the stellar radius $R$ and its gravitational mass $M=m(R)$.

\begin{figure}[h]
\begin{center}
\includegraphics[width=0.48 \textwidth]{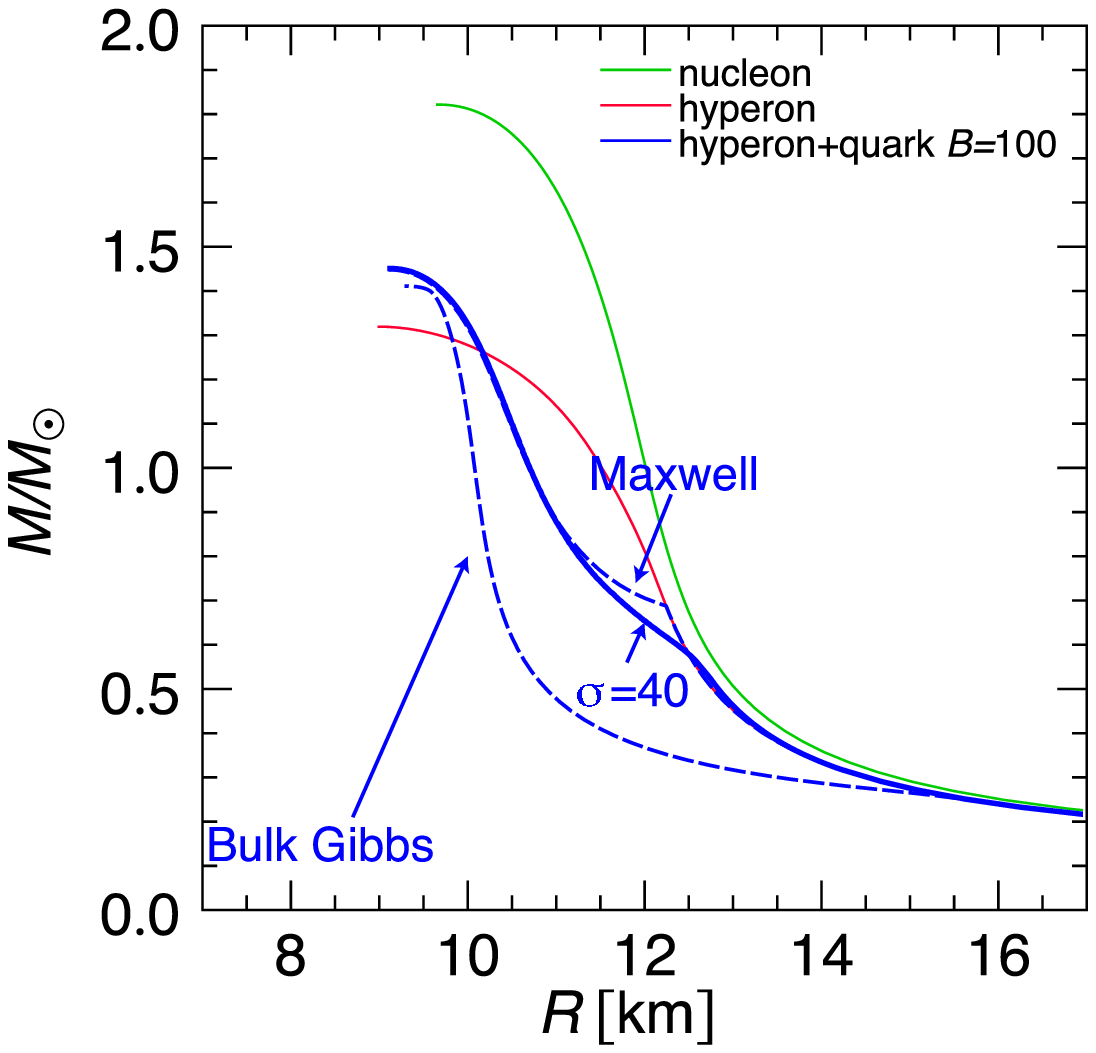}
\end{center}
\caption{
Mass-radius relation of calculated neutron stars. Green curve indicates the result with only nucleons.Red curve is that with hyperons. Blue curves are results with hadron matter and quark matter by different treatments of the mixed phase
Note that the Maxwell construction is incorrect in the present case, but it is still useful as an eye guideline.
}
\label{figMRrelation}
\end{figure}

We show the effect of the HQ phase transition on
the relation between the EOS and bulk structures of compact stars
in Fig. \ref{figMRrelation}. The left panel is the EOSs and the right panel
 shows the mass-radius relation of calculated neutron stars.
As for the EOS, we shows three types of the HQ phase transitions;
 c.g. the EOS by the bulk Gibbs calculation, the Maxwell construction, and
full calculation considering the finite-size effects.
Here, the value of the surface tension is set as $\sigma=40$ MeV$\cdot$fm$^{-2}$.
We also show the uniform hadron and quark matter for comparison.
Using them, we can calculate the mass-radius relations shown in the right panel.
It is well-known that the typical masses of neutron stars are distributed
 in a narrow band close to 1.4 times the solar mass $M_\odot$ \cite{nsmass}.
The maximum mass of neutron star is enough larger than these masses.
With inclusion of hyperon degree of freedom, however,
the softening of matter reduces the possible mass of neutron stars
as low as 1.3 $M_\odot$, which contradicts with the most observations.
The blue curves show the cases with the HQ mixed phase.
The upper most one and the bottom are the results with the EOS obtained
by the Maxwell construction and the bulk Gibbs calculation, respectively.
The result with inclusion of the full pasta structures
with $\sigma=40$ $\rm MeV \cdot fm^{-2}$ lies just below the Maxwell construction case.

In any case with the HQ mixed phase, the calculated maximum mass
is slightly above $1.4 M_\odot$, which might be tolerable in confrontation with the observation.
What is important is that the mixture of hyperons in nuclear matter
softens the EOS, which causes the contradiction with the observation,
i.e.\ the ``maximum-mass problem''.
By considering the deconfinement transition to quark matter, the EOS gets stiffer at
higher density region.
This stiffness at high density sustains the neutron stars with $M\sim 1.4 M_\odot$.

Recently Demorest et.al. have reported that there is a two-solar-mass neutron star through the measurement of the Shapiro delay~\cite{dem10}. Motivated by this observation, some authors studied the possibility of hard EOS considering exotic matter; c.g. Ref.\cite{Weissenborn11} for hadron matter with hyperons, and Ref.\cite{masuda12} the matter including quarks. We are also planning to update our EOS models to other models to satisfy the observation.

Another possibility is the observations of gravitational waves radiated from neutron stars or the observations of stellar oscillations itself, where the observed spectra can reveal the interior properties \cite{AK1996,AK1998}. This technique is well-known as ``gravitational-wave asteroseismology'' or just ``asteroseismology,'' which is similar to the seismology in the Earth and the helioseismology in the Sun. In this section, we focus on the possibility to see the properties of the HQ mixed phase via the observations of gravitational waves in subsection \ref{sec:sotani-1} and stellar oscillations in subsection \ref{sec:sotani-2}.

In the last part of this section, we introduce another possibility to know the effects of the HQ mixed phase on compact-star phenomena; thermal cooling of neutron stars.
A cooling theory of neutron stars has been well-discussed but a still uncertain issue~\cite{st98,bk09}.
Cooling depends on the internal state which is in the range of ultrahigh density around/above the normal density. It may be directly connected to the EOS discussed in the previous section.

\subsection{Gravitational Waves}
\label{sec:sotani-1}

The gravitational waves are oscillations of spacetime itself. Due to their strong permeability, the gravitational waves can carry away the raw information about the wave sources. So, observing the gravitational waves emitted from neutron stars, it could be possible to collect the astronomical data \cite{AK1996,AK1998,KAA2001}, to reveal the properties of dense matter \cite{STM2001,SH2003,SKH2004,GK2009,knippel09,SYMT2011}, and to probe the gravitational theory in the strong gravitational field \cite{SK2004,SK2005,S2009a,S2009b,S2011a}. Now, in order to directly detect the gravitational waves, several ground-based detectors are in operation, and the projects to build the next generation detectors are also moving ahead \cite{B2005,LISA,DECIGO}, which could enable us to observe the gravitational waves in the not-so-distant future.

Considering the gravitational waves radiated from the spherically symmetric stars, their oscillations can be classified into two families with their parities, i.e., odd (axial) and even (polar) parities. Since the odd parity oscillations are completely decoupled with the even parity oscillations, one can examine each type of oscillations separately. The odd parity oscillations are incompressible motion, while the even parity oscillations involve the deviation of density. Thus, from the observational point of view, the even parity oscillations might be important in the emission of gravitational waves. Here, in order to determine the specific oscillation frequencies in neutron stars, we adopt the relativistic Cowling approximation, where the perturbation of gravitational potential should be neglected. Namely, the fluid would oscillate on a fixed background spacetime. In practice, even with this simple approximation, one can qualitatively see the features for oscillation frequencies of emitted gravitational waves.

\begin{figure}[t]
\begin{center}
\begin{tabular}{cc}
\includegraphics[scale=0.45]{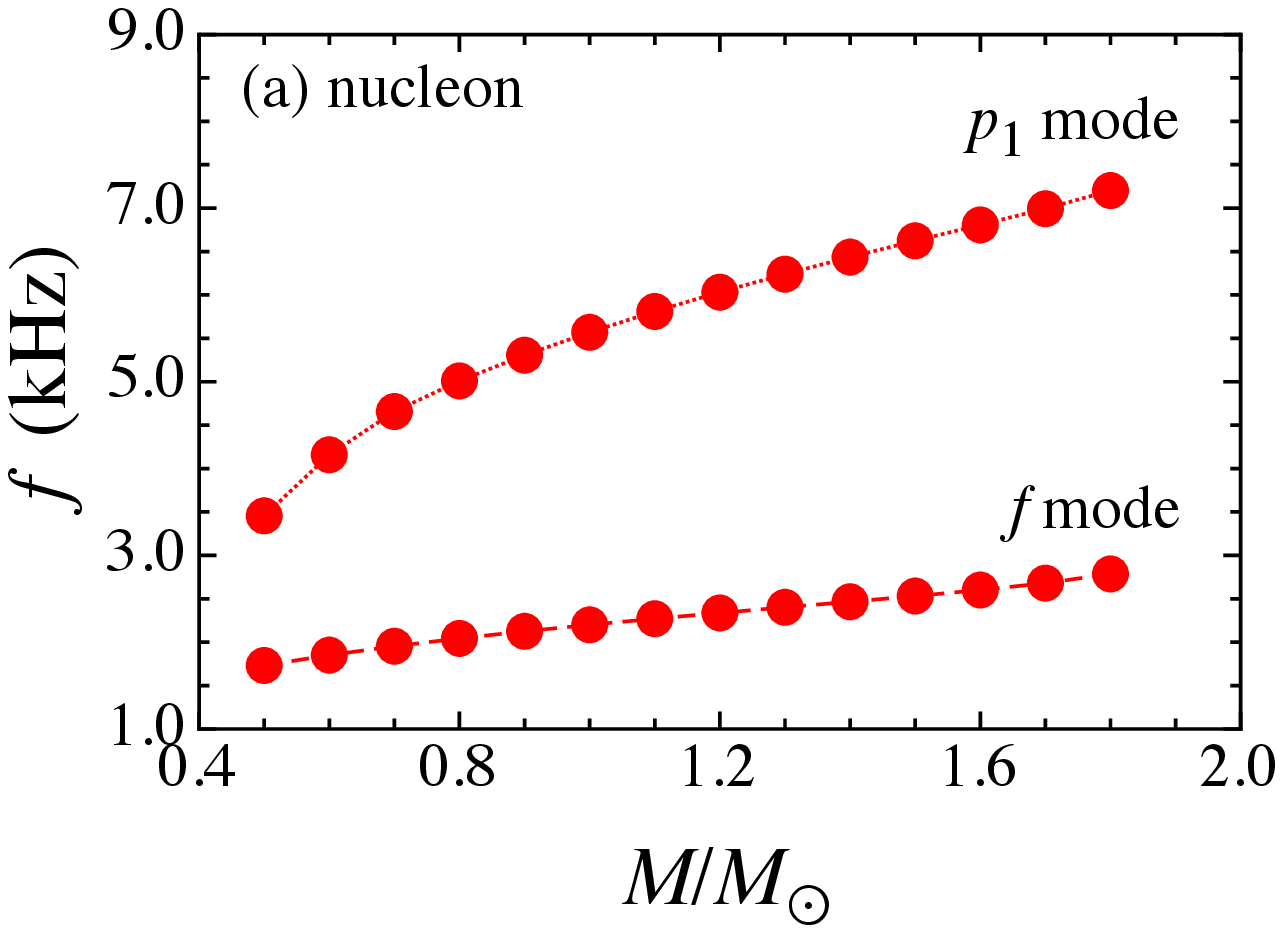} &
\includegraphics[scale=0.45]{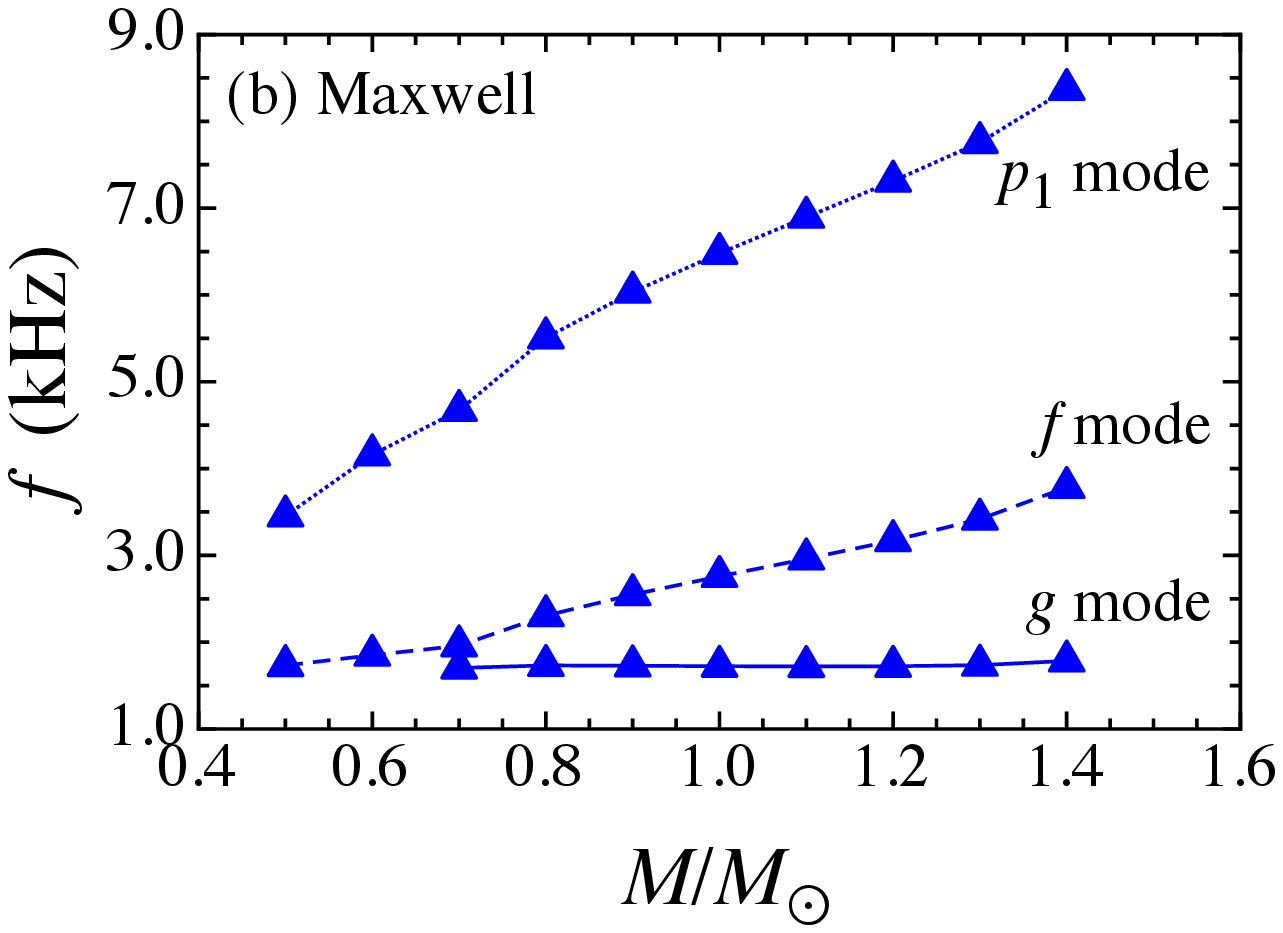} \\
\end{tabular}
\end{center}
\caption{
The frequencies of $f$ and $p_1$ modes for the stellar models with (a) nucleon and (b) the HQ mixed phase adopted the Maxwell construction as a function of the stellar mass. In the right panel, the additional frequencies of $g$ modes are also shown as well as the frequencies of $f$ and $p_1$ \cite{SYMT2011}.
}
\label{fig:spectrum}
\end{figure}

\begin{figure}[t]
\begin{center}
\includegraphics[scale=0.45]{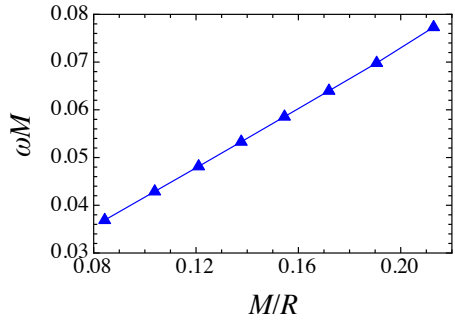}
\end{center}
\caption{
The normalized eigenfrequencies of $g$ modes with EOSs considering the Maxwell construction as a function of the stellar compactness $M/R$ \cite{SYMT2011}.
}
\label{fig:gmode}
\end{figure}

During the stellar oscillations, many kinds of gravitational waves could be radiated. The fundamental ($f$) and the pressure ($p$) modes can be excited in the spherically symmetric stars without the density discontinuity inside the star, while the gravity ($g$) modes can be also excited with the density discontinuity. As shown in Fig. \ref{fig:spectrum}, the stellar models with the EOSs adopted in this article except for the EOS with the Maxwell construction have only $f$ and $p$ modes, where the details how to determine the specific oscillation frequencies can be seen in Ref. \cite{SYMT2011}. On the other hand, in the stellar models constructed with the EOS adopted the Maxwell construction whose masses are more than $0.7M_\odot$, the $g$ mode is also excited as well as $f$ and $p$ modes. It should be noticed that since the stellar models with $M<0.7M_\odot$ do not have density discontinuity inside the star because of the small central density, the $g$ mode can not be excited. Since, as mentioned before, the $g$ mode can be excited due to the existence of density discontinuity inside the star, the observations of the $g$ mode can tell us such an information. In fact, the typical frequencies of $g$ modes are in the range from a few hundreds Hz up to kHz, it could be possible to detect via the ground-based gravitational wave detectors. Additionally, as shown in Fig. \ref{fig:spectrum}, the frequencies of $g$ mode for the stellar models adopted in this article are almost independent of the stellar mass, i.e.,  $\sim 1.73$ Hz. While, as shown in Fig. \ref{fig:gmode}, it is found that the $g$ mode frequencies can be  well expressed as a function of stellar compactness, such as $\omega M=0.3130(M/R)+0.0103$. That is, via the detailed observations of $g$ mode gravitational waves, one could know the stellar properties with the help of the additional observations about the stellar parameters such as the mass.

\begin{figure}[t]
\begin{center}
\begin{tabular}{cc}
\includegraphics[scale=0.45]{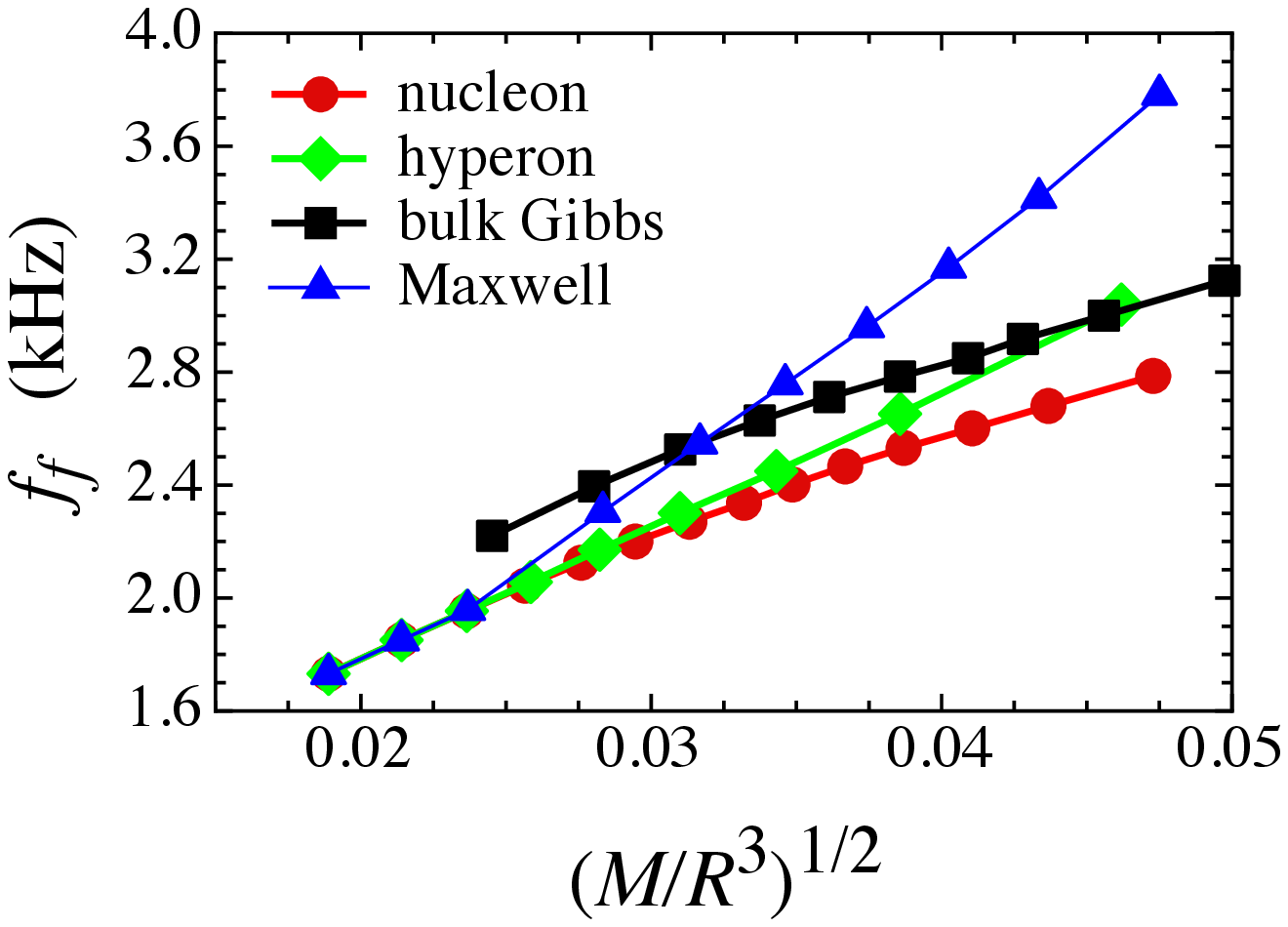} &
\includegraphics[scale=0.45]{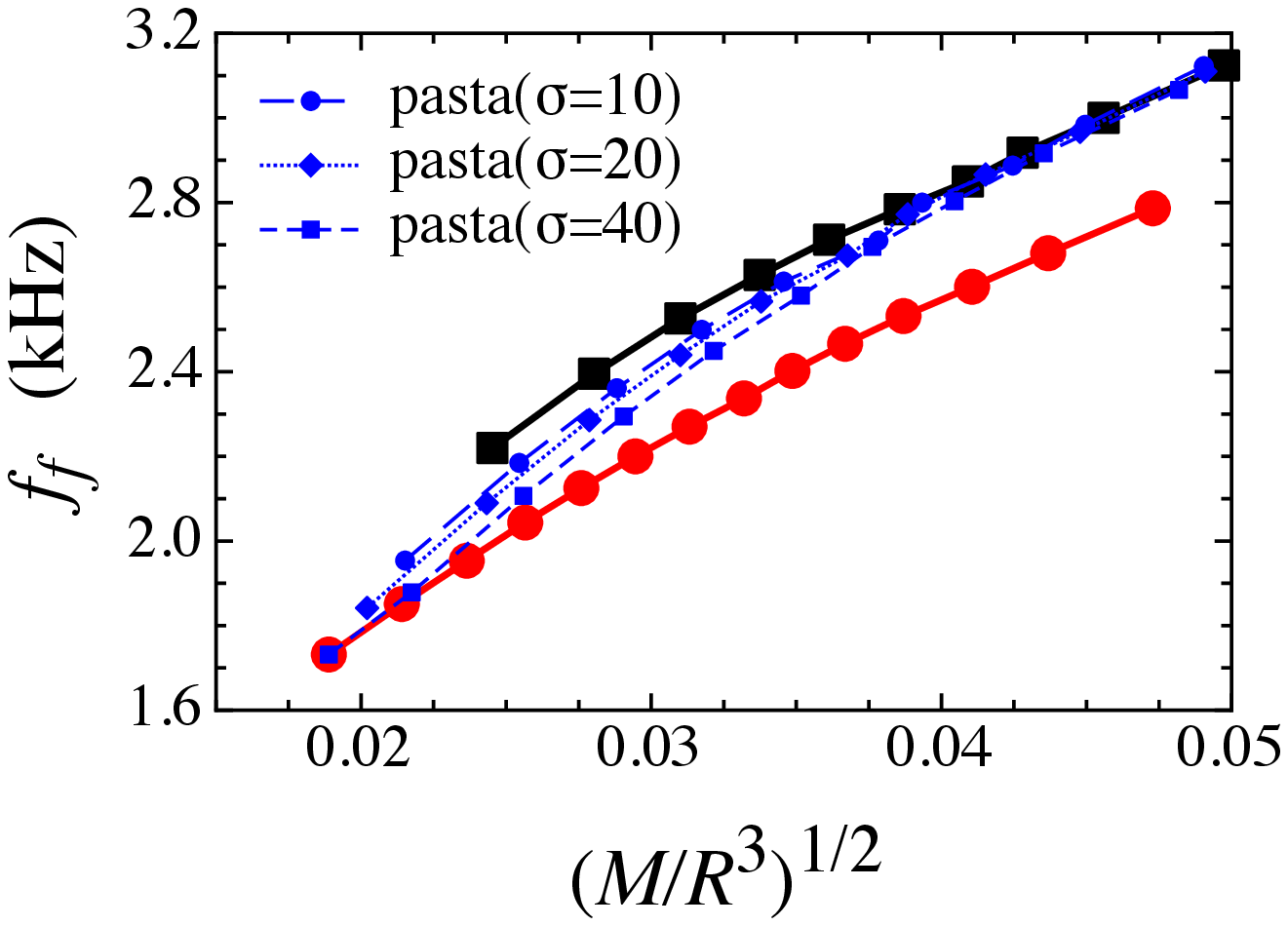} \\
\end{tabular}
\end{center}
\caption{
The frequencies of $f$ modes for the stellar models with different EOSs as a function of the stellar average density $(M/R^3)^{1/2}$. The results with nucleon, hyperon, HQ mixed phase adopted the bulk Gibbs conditions~(GCN) or the Maxwell construction are shown in the lift panel, while the right panel focuses on the results with the EOS considered the finite-size effects. Here the units of each $\sigma$ are MeV $\cdot$ fm$^{-2}$ \cite{SYMT2011}.
}
\label{fig:fav}
\end{figure}

\begin{figure}[t]
\begin{center}
\begin{tabular}{cc}
\includegraphics[scale=0.45]{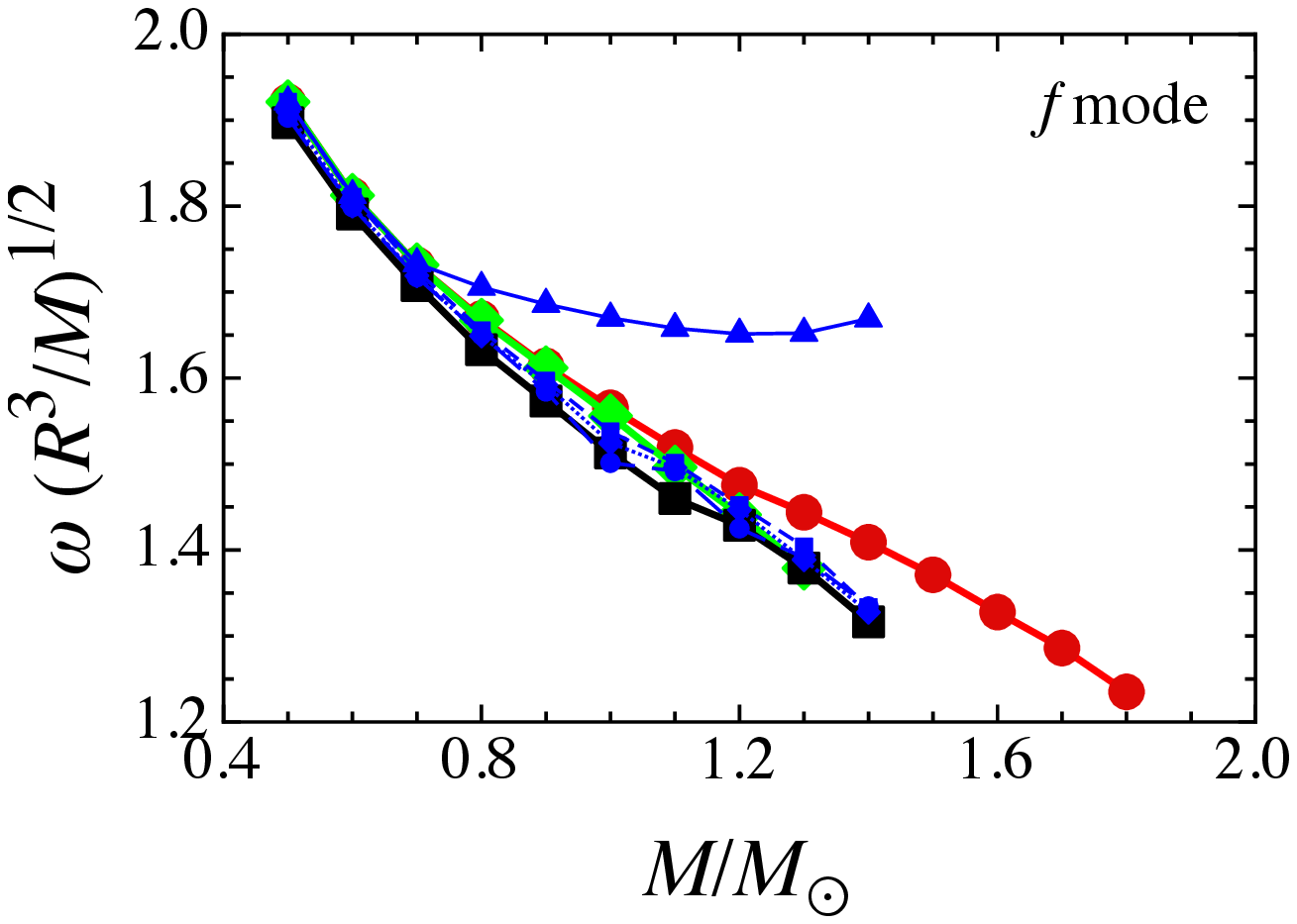} &
\includegraphics[scale=0.45]{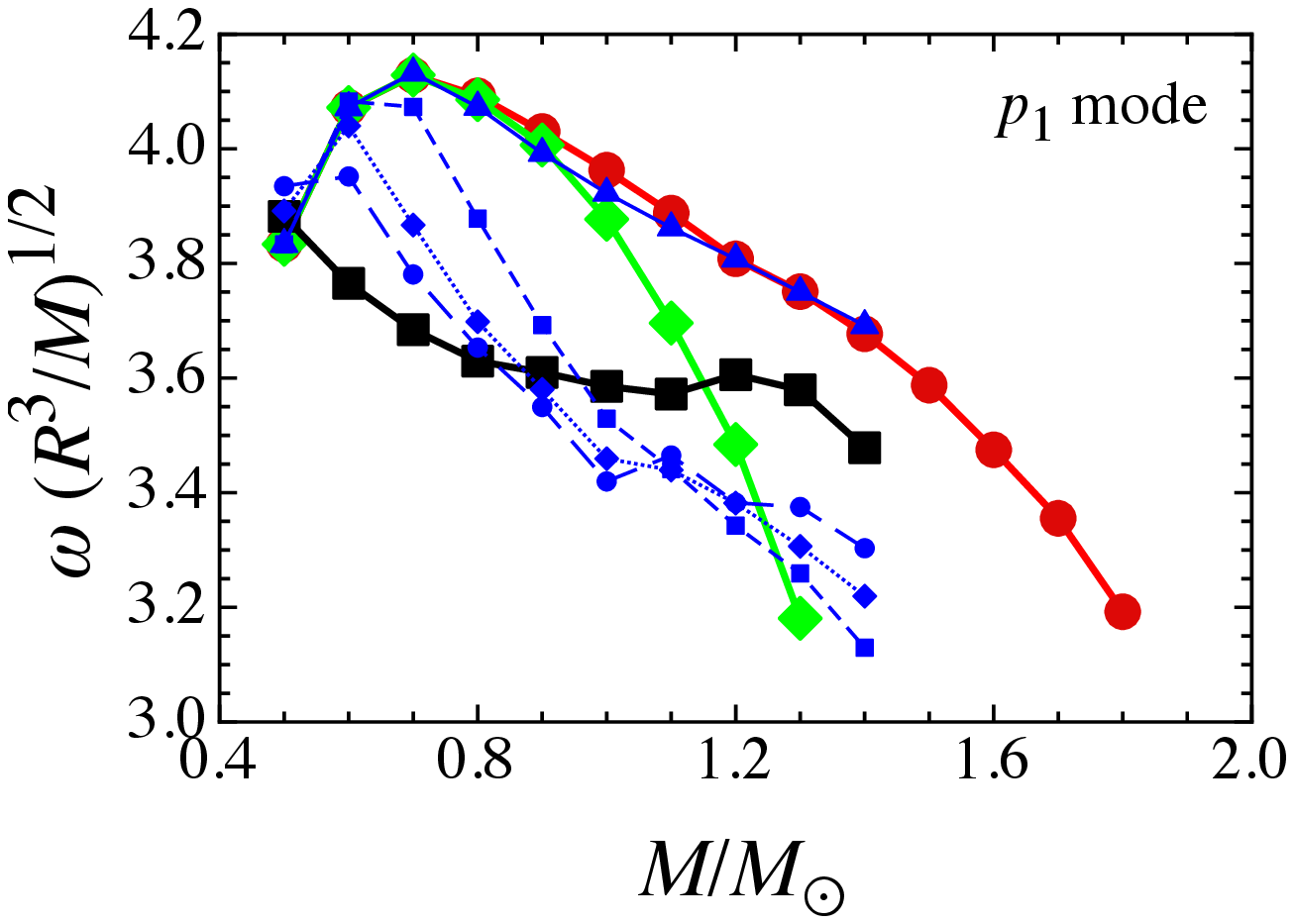} \\
\end{tabular}
\end{center}
\caption{
The normalized eigenfrequencies of $f$ modes (left panel) and $p_1$ mode (right panel) as a function of the stellar masses. The marks in the figures are the same meaning as in Fig. \ref{fig:fav} \cite{SYMT2011}.
}
\label{fig:normalizedf}
\end{figure}

 Furthermore, it is well-known that the $f$ mode frequency can be associated with the stellar average density defined as $(M/R^3)^{1/2}$ \cite{AK1996,AK1998}. This could be physically explained by considering the relation between the sound speed and the propagation time of the fluid perturbation inside the star. The $f$ mode frequencies  calculated for the stellar models with the adopted EOSs are shown in Fig. \ref{fig:fav}. Roughly speaking, one can observe from this figure that the frequencies for the stellar models without the density discontinuity, i.e., except for the EOS with the Maxwell construction, are almost similar behaviors. Thus, it might be difficult to reveal the properties of the HQ mixed phase via the observations of $f$ mode gravitational waves. However, it should be emphasized that the $f$ mode frequencies for the stellar model with the density discontinuity (with EOS adopted the Maxwell construction) can behave qualitatively different from the others. That is, via the observations of not only $g$ modes but also $f$ modes gravitational waves, one might be able to prove the existence of the density discontinuity inside the star. Furthermore, from Fig. \ref{fig:normalizedf} for the normalized eigenfrequencies of $f$ and $p_1$ modes, one can obviously see the difference between the frequencies of $f$ mode gravitational waves with and without density discontinuity as mentioned the above, while the  $p_1$ mode gravitational waves seem to depend on the adopted EOSs even if the density discontinuity does not exist inside the star. Therefore, the detailed observations of $p$ mode gravitational waves could reveal the properties of the HQ mixed phase.

\subsection{Shear Oscillations}
\label{sec:sotani-2}

Unlike the gravitational waves emitted from neutron stars, there exist the observed evidences of the stellar oscillations, i.e., the quasi-periodic oscillations (QPOs) in giant flares radiated from soft-gamma repeaters (SGRs). Up to now, three giant flares have been observed, such as SGR 0526-66 in 1979, SGR 1900+14 in 1998, and SGR 1806-20 in 2004, and the existence of a lot of QPO frequencies in the afterglow of giant flares are found via timing analysis, whose frequencies are in the range of tens Hz up to kHz \cite{WS2006}. In order to theoretically explain the observed QPO frequencies, many examinations have been done in terms of the torsional oscillations in the crustal region and/or the magnetic oscillations (e.g., \cite{L2006,L2007,SA2007,SKS2007,SKS2008,SCK2008,SK2009,GCFMS2011,CK2011}). If the observed QPO frequencies are associated with the torsional oscillations in neutron star crust, one might be able to obtain the information about the crust properties \cite{pethick98,SW2009,S2011b,GNHL2011,SNIO2012}. The torsional oscillations are incompressible, i.e., odd parity oscillations, where the elasticity plays a role as a restoring force, and such elasticity can be characterized by the shear modulus $\mu$ \cite{ST1983}. Although the derivation of $\mu$ for the complex structure have never been done, the expression of $\mu$ due to the Coulomb interaction in the body center cubic (bcc) in neutron star crust is known as
\begin{equation}
 \mu = 0.1194 n_i(Ze)^2/a, \label{eq:shear}
\end{equation}
where $n_i$, $Ze$, and $a$ are ion number density, charge of ion, and average ion spacing defined as $4\pi a^3/3=1/n_i$, respectively \cite{SHOII1991}. Note that, averaging the all directions and assuming the zero temperature, this expression is derived   from the Monte Carlo calculations.

On the other hand, as mentioned the above, the possibility to appear the nonuniform structure (the HQ pasta phase) in the high density region inside the neutron stars is suggested \cite{hyp07,yas09}, which is similar structure to the pasta phase in the neutron star crust \cite{Lorenz1993,Oyamatsu1993}. If this suggestion is true, similar to the neutron star crust, one might be able to consider the shear in the HQ mixed phase, and the torsional oscillations could be excited even in the HQ mixed phase~\cite{owen11,SMT2012}. In order to examine such torsional oscillations, for simplicity, we consider the shear only for the quark spherical droplet in the uniform hadronic matter, although the different nonuniform structures could appear in the HQ pasta phase, such as cylindrical and slab structures. Due to this assumption, the shear in the HQ mixed phase effectively makes an underestimation. So, the calculated frequencies of torsional oscillations could be corresponding to the lower limit, because the frequencies of torsional oscillations can be proportional to the shear speed defined as $v_s=(\mu/\rho)^{1/2}$. Moreover, to adopt the expression for $\mu$, Eq. (\ref{eq:shear}), we consider that $n_i$ should be the number density of quark spherical droplet in the hadronic sea, while $Ze$ should be the total charge included in the quark spherical droplet. As a typical value of the surface tension $\sigma$, we adopt three values, such as $\sigma=10$, 20, and 40 MeV $\cdot$ fm$^{-2}$. The detail for calculations can be seen in \cite{SMT2012}.

\begin{figure}[t]
\begin{center}
\includegraphics[scale=0.45]{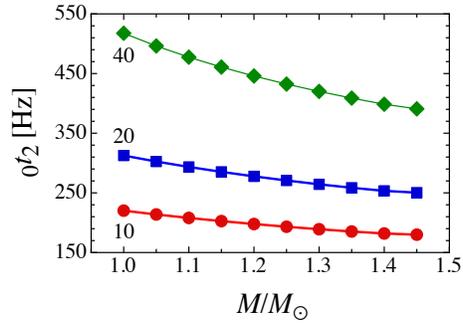}
\end{center}
\caption{
With $\sigma=10$, 20, and 40 MeV$\cdot$ fm$^{-2}$, the frequencies of the $\ell=2$ fundamental torsional oscillations ${}_0t_2$ are shown as a function of stellar mass $M/M_\odot$  \cite{SMT2012}.
}
\label{fig:torsional}
\end{figure}

\begin{figure}[t]
\begin{center}
\includegraphics[scale=0.45]{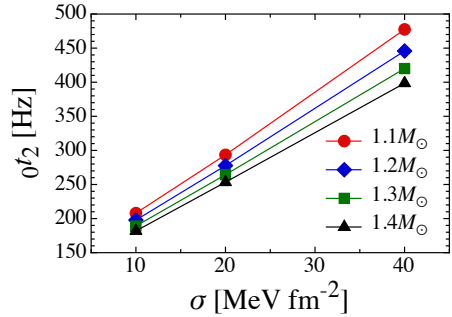}
\end{center}
\caption{
Frequencies of fundamental torsional oscillations with $\ell=2$ as a function of surface tension with different values of stellar mass \cite{SMT2012}.
}
\label{fig:t2-s}
\end{figure}

The calculated frequencies of the $\ell=2$ fundamental torsional oscillations in the HQ mixed phase, ${}_0t_2$, are shown in Fig. \ref{fig:torsional} as a function of the stellar mass. Since the frequencies of fundamental torsional oscillations in the neutron star crust are around tens Hz \cite{SKS2007,ST1983}, one can observe around 10 times larger frequencies in the HQ mixed phase. This is why the shear modulus in the mixed phase calculated from Eq. (\ref{eq:shear}) becomes about $10^3$ times larger than that in crust. As a result, the propagation time with the shear speed becomes about 10 times smaller than that in the crust. Then, one can estimate that the frequencies of torsional oscillations in the mixed phase could become roughly 10 times as large as those in crust region. Additionally, one can see from Fig. \ref{fig:torsional} that the frequencies of fundamental torsional oscillations depend strongly on the value of $\sigma$. In practice, the frequencies for $\sigma=20$ and 40 MeV$\cdot$ fm$^{-2}$ are $\sim$40\% and $\sim$120 \% larger than those for $\sigma=10$ MeV$\cdot$ fm$^{-2}$. Thus, if one would identify the observed frequencies as the torsional oscillations in the HQ mixed phase, one can probe the properties of such exotic structure.
Additionally, with the fixed stellar mass, we plot the frequencies of fundamental torsional oscillation with $\ell=2$ in the HQ mixed phase as a function of surface tension $\sigma$ in Fig. \ref{fig:t2-s}. From this figure, one can obviously observe the linear relation between the frequencies and surface tension. Thus, with the help of the other observation of stellar mass, one can make a constraint in the value of $\sigma$ via the observation of the frequencies of fundamental torsional oscillations in the HQ mixed phase. Now, we should remark the possibility of the detection of this type oscillation. In fact, if the star is exactly spherically symmetric, one might have no chance to observe this oscillation, because the HQ mixed phase is surrounded by the hadronic fluid matter and the oscillation is completely confined in the HQ mixed phase. But, considering the realistic stellar model, there exist the rotational and/or magnetic effects, which make a deformation of stellar shape. Under this situation, the torsional oscillations with axial parity can be coupled with the even parity oscillations mentioned in Sec. \ref{sec:sotani-1}. Probably, this is one opportunity to observe the torsional oscillation in the HQ mixed phase. Anyway, the examinations in the more complicated system will be done somewhere.
Furthermore, since the resulting frequencies of fundamental torsional oscillations are order of 100 Hz, some of the QPO frequencies observed in giant flares, for example 150 Hz and even 626.5 Hz  in SGR 1806-20 and 155 Hz in SGR 1900+14 \cite{WS2006}, might be associated with the torsional oscillations in the HQ mixed phase.

\subsection{Cooling of Neutron Stars with the Hadron-quark Mixed Phase}
\label{sec:noda-1}

Now we focus on the cooling theory of isolated neutron stars.
Isolated neutron stars do not have internal heat source (someone says magnet-breaking heating affects to the cooling, but the effect is not so large), and they are born
at the sites of supernovae explosion where the temperature rises to 30-50 MeV in the core of PNS as we described in the last section. Isolated neutron stars evolve thermally by emitting energy and losing their temperature immediately, as a result the temperature becomes to a few MeV (or $\sim10^{10}\mathrm{K}$) in a minute\cite{burrows, fischer}. In this subsection we discuss the cooling of neutron stars after a few minutes from their birth. Hence initial temperature is assumed as $\sim 1$ MeV in our calculations.

The most efficient energy-release process in neutron stars is the neutrino emission, based on neutron beta-decay.
The well-known process named direct URCA process is combination of neutron beta-decay and inverse reaction.
\be
	n \to p + e^- + \bar{\nu}_e, \ \ \  p + e^-  \to n + \nu_e
\ee
Working forward and inverse (electron capture) process, electron neutrinos and anti-electron neutrinos are emitted and carry some energy.
Emitted neutrinos travel through the stars without any interactions due to the large mean free path.
However, due to the abundances of proton and neutron, direct URCA process is  kinematically forbidden in the neutron stars~\cite{sha}.
To solve this issue, the modified URCA process was promoted, which involves another neutron to conserve momentum.
\be
	n + n \to n + p + e^- + \bar{\nu}_e, \ \ \ n + p + e^- \to n + n + \nu_e
\ee
The modified URCA process has a lower emissivity; it is about five orders of magnitude weaker than one with the direct URCA.

Other processes also work on neutron stars cooling, such as Bremsstrahlung (neutrino emission) and photon emission, but they work dominantly in the crust (Bremsstrahlung), or the surface (photon emission) of neutron stars, and their emissivities are weaker than the core processes.

Modified URCA and other processes could explain most of neutron stars observational data, with some exception.
Therefore the ``Exotic (Non-Standard) Cooling'' which works in some kinds of exotic phases, is introduced and the ordinary cooling is called the ``Standard Cooling''~\cite{fri79}.
The standard cooling works in all the neutron stars, and the exotic cooling works in some of neutron stars.
The exotic cooling works above the threshold density, due to some kind of exotic situations where the exotic phase appears.
The exotic situation can be said as pion condensation, kaon condensation, existence of hyperons or quark matter~\cite{sha}.
In the exotic state, the neutrino emission process plays as the direct URCA, due to the degrees of freedom.
Once the exotic cooling process works, star cools faster than standard one~(eg. \cite{yak05,gus05}).

Stars with the exotic cooling have larger central density than ones with the standard cooling.
The mass of stable neutron stars increases with the central density, therefore stars with the exotic cooling have larger mass than standard ones~\cite{lat91}.
Also, the neutrino emissivities of these processes including the modified URCA have density dependent; it is monotonically increasing with the density.
It is believed that lighter neutron stars cools slower and heavier one cools faster.

The matter in neutron stars is well degenerated; neutrons and protons can create the Cooper pairs and condensate to superfluid (superconducting for protons) state at some temperature.
The transition between normal to superfluid state, the latent heat escapes by causing neutrino pair creation~\cite{yak04}.
This process is called as PBF (Pair Breaking and Formation) process, and it works only where the temperature of the region is near to the critical temperature of superfluidity. In neutron stars, there are three possible nucleon superfluid states (${}^1S_0$ and ${}^3P_2$ for neutrons, and ${}^1S_0$ for protons), and they might have three PBF working regions. The neutron ${}^3P_2$ state works at the highest density of these three, and it may become the dominant cooling process at specific density and time~\cite{yak99,gus02}. The transition does not last long time and occurs just once in a density region, therefore the integrated effect is limited.
However, once matter becomes to superfluid state, thermodynamical properties decreases and the emissivity of neutrino also decreased.
Nucleon superfluidity works both of acceleration and breaking of cooling.
The neutrino emission by the transition to the superfluid state works all the neutron stars, and called as ``minimal cooling'' including the modified URCA, Bremsstrahlung and PBF~\cite{pg04}.

There are some cooled stars whose effective temperature cannot be explained
by the standard cooling scenario, which does not include the neutrino emission caused by nucleon superfluidity.
It needs stronger cooling process as in the case of J0205+6449 in 3C58 (hereafter, ``3C58'') or Vela pulsar (B0833-45). Also an accreting neutron star SAX J1808 requires strong cooling. 3C58 and Vela may be explained by the minimal cooling model which includes nucleon superfluidity~\cite{pg09}.
However, SAX J1808 needs stronger cooling than the minimal cooling \cite{heinke08}.

In recent observation, two important data for cooling of neutron stars are observed and analyzed.
One is the central source of Cassiopeia A (hereafter ``Cas A'')~\cite{ho09,ho10} and the other is J1614-2230~\cite{dem10}.
They give strong constraints on the cooling of neutron stars.
J1614-2230 is discussed in the beginning of this chapter, we introduce about the Cas A issue here.

Cas A is the youngest-known supernova remnant in our Milky Way and it locates $\sim3.4~\mathrm{kpc}$ distant from the solar system~\cite{rd95}. The supernova explosion occurred about 330 years ago, but due to absorption by interstellar medium, there are no exact historical records except for
unclear detection by J. Flamsteed in 1680~\cite{aw80}. Recently,
 \cite{ho09}  and \cite{ho10} have analyzed the X-ray spectra of Cas A. They
give the effective temperature and possible regions occupied by the mass-radius relations. Since Cas A is the isolated remnant, uncertainty of the mass-radius relation could be large.
The lowest mass obtained from $\chi^2$ fitting is about $1.5M_\odot$. Considering the age of $t=330$~yr, the effective temperature $T_{\rm eff}$ of Cas A must occupy a point of a cooling curve
due to the standard cooling
scenario on the $(T_{\rm eff}-t)$ plane. This gives strong constraint on the EOS and cooling processes.
Furthermore, \cite{ho10} reported the observation of $T_\mathrm{eff}$ for Cas A in recent 10 years. \cite{yak11}, \cite{pg11} and \cite{sht11} insist that the rapid decrease in $T_\mathrm{eff}$ during the years shows that the transition to nucleon superfluidity occurs.

\subsubsection{Cooling of Hybrid Stars with the Hadron-quark Mixed Phase}

Considering the HQ mixed phase for the EOS of neutron star, it is important to apply for the cooling theory.
The quark matter is known as an exotic matter which causes strong neutrino emission.
Therefore, once the quark phase appears, strong neutrino emission cools the star quickly.
The neutrino emissivity is calculated as~\cite{iwa80}
\be
	\varepsilon_{\nu, q}=8.8\times10^{26}\alpha_s\frac{n_B}{n_0}Y_e^{1/3}T_9^6   \, \, \ \rm erg~s^{-1}~cm^{-3}. \label{eq:q-emi}
\ee
This emissivity can work inside of quark matter, and we show the following assumption.

We constructed a model which includes the HQ mixed phase. Considering the
first order phase transition between the hadron and quark phases, it would be plausible that both phases coexist and form some kinds of the mixed phase. It has been shown that the mixed phase could form geometrical structures~\cite{hyp07}; \cite{yas09} have made EOS of mixed-phase under a Wigner-Seitz (hereafter ``WS'') approximation using the MIT Bag model for the quark phase at finite temperature.

We employ an EOS with the same framework using the bag constant
$B=100~ \mathrm{MeV~fm^{-3}}$, the coupling constant $\alpha_\mathrm{S}=0.2$, and the surface tension parameter $\sigma = 40~\mathrm{MeV~fm^{-2}}$.
For the hadron phase, we adopt results of the BHF theory including hyperons, $\Lambda$, and $\Sigma^-$~\cite{baldo98a,baldo98b,schulze95,vidana95}. Since the BHF results is inappropriate for low-density matter in
the crust, we apply  EOS of BPS~\cite{bps71} for the crust. The EOS gives the maximum mass $1.53M_\odot$ with the radius $8.6~ \mathrm{km}$, and the mass lies within the limits of the observation of Cas A.

\begin{figure}[htbp]
	\begin{center}
		\includegraphics[width=0.5\linewidth,keepaspectratio,clip]{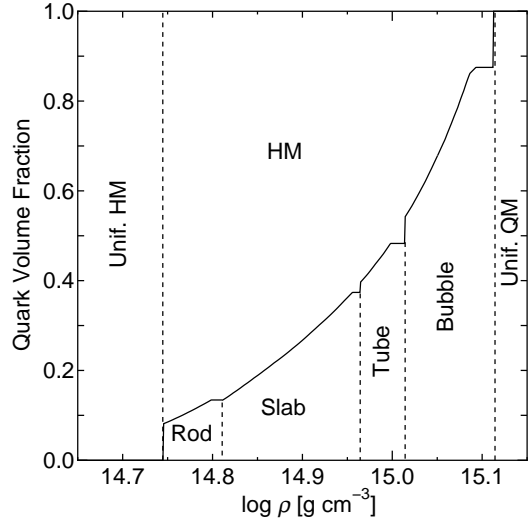}
	\end{center}
	\caption{Volume fractions of quark matter phase having particular geometrical structures with a bag constant $B=100~\mathrm{MeV}~\mathrm{fm}^{-3}$ and a coupling constant
 $\alpha_\mathrm{s} = 0.2$~\cite{hyp07,yas09}.}
	\label{fig:MP}
\end{figure}

Using the WS approximation, we obtain a cell radius of each phase, and calculate the volume fraction of quark matter in the mixed phase as seen in Fig. \ref{fig:MP}.

It is difficult to calculate the neutrino emissivity in the mixed phase directly.
Therefore, the volume fraction $F$ is multiplied to the original quark neutrino emissivity $\varepsilon_{\nu,q}$ of (\ref{eq:q-emi}); The total emissivity by quarks is set to be $\varepsilon_\nu = F \varepsilon_{\nu,q}$.
We adopt the well-known standard cooling process for hadronic matter~\cite{fri79} : Modified URCA process for the higher density region and Bremsstrahlung process for the crust.

\begin{figure}[htbp]
	\begin{center}
		\includegraphics[width=0.5\linewidth,keepaspectratio,clip]{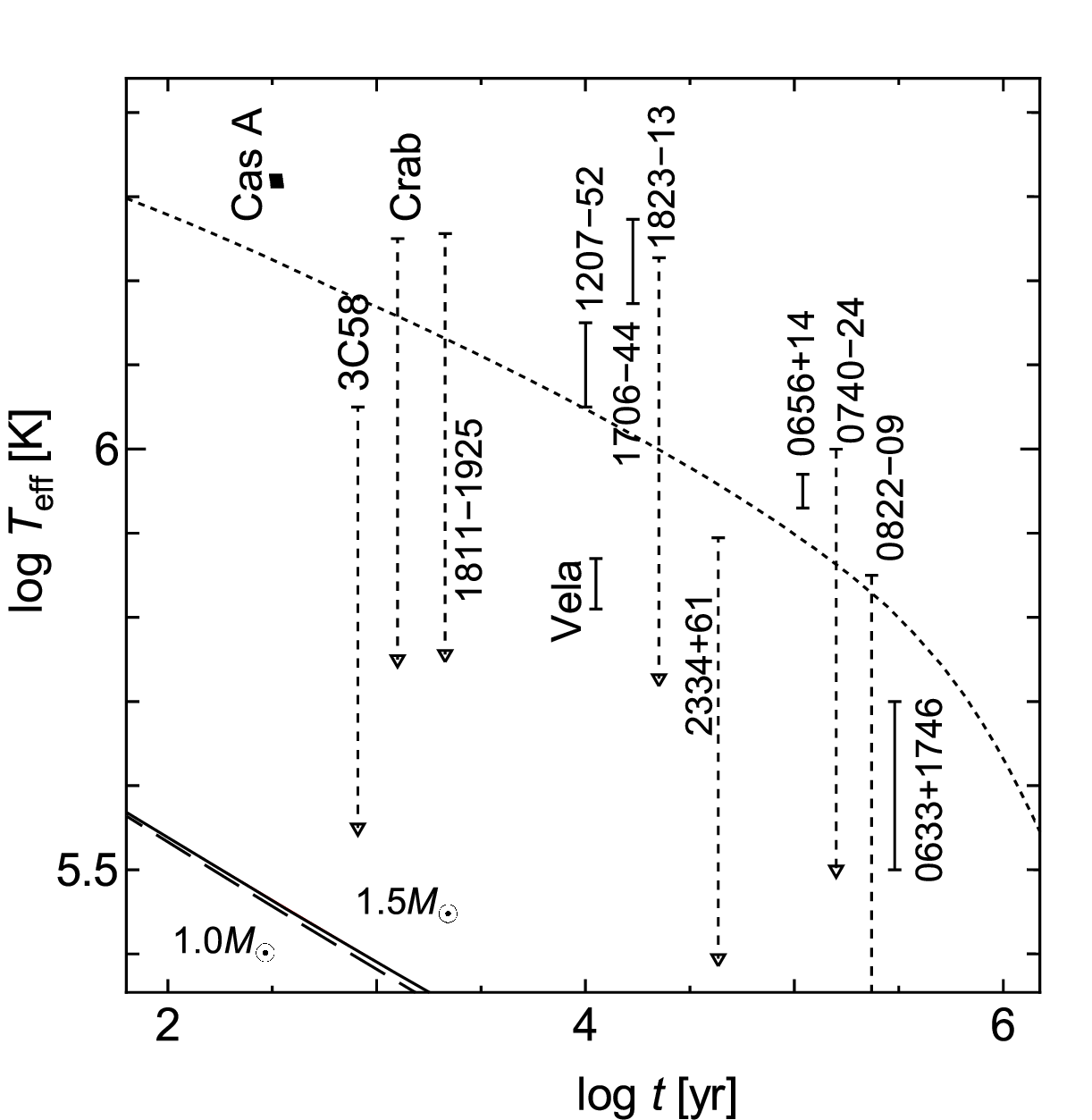}
	\end{center}
	\caption{Cooling curves with the HQ mixed phase. Solid and dashed curves denote cooling curves of the hybrid stars with the HQ mixed phase, and the dotted curve denotes the cooling curve of the neutron star with standard cooling. The curves show the models of hybrid stars with the HQ mixed phase become too cold to explain observed data.}
	\label{fig:CC}
\end{figure}

Considering the HQ mixed phase, it becomes possible to reduce the neutrino emissivity by quarks.
However, the cooling rate is still too strong to explain cooled stars with lower limit (such as Vela), it requires some kind of reduction of neutrino emissivity.

\subsubsection{Cooling of Hybrid Stars with the Hadron-quark Mixed Phase and Color Superconductivity}

Fortunately, there is possible factor of reduction of neutrino emissivity in the quark phase.
It is the color superconducting (hereafter ``CSC'') phase of quarks~\cite{bl00}.
There are some sorts of quark pairings such as CFL or 2SC according to the degrees of freedom of quark flavor and color.
In the CFL phase, all colors and flavors quarks have a common Fermi momenta, and all quarks create pairs and condensated.
Meanwhile $u$ and $d$ quarks of two colors have same Fermi momenta in the 2SC  phase, and some quarks are unable to create pairs~\cite{alf08}.
The CFL phase is likely to appear at higher density than the 2SC phase~\cite{rus05}.
It is considered that the energy gap of $\Delta \sim 10~\mathrm{MeV}$ is very large compared with the temperature of the center of compact stars , $T_\mathrm{C} \sim \mathrm{keV}$,(eg.~\cite{shm10}).
Once matter becomes superconducting, neutrino emissivity must be suppressed similar to the hadronic superfluidity~\cite{hess11}. Due to the large energy
gap and it could be proportional to $\exp\left(-\Delta/k_\mathrm{B}T\right)$, where $T$ is the temperature at the relevant layer and $k_{\rm B}$ is the Boltzmann constant.
Therefore, in the color superconducting phase with a large energy gap ($\Delta \gg T$), neutrino emissivity by quarks is almost negligible~\cite{alf08}.
We note that we do not need to consider which kinds of CSC pairing appears, if we adopt a large enough energy gap.

We assume that the energy gap in the color superconducting phase is very large and the phase appears in the
density region above the threshold density $\rho_\mathrm{T}$. For simplicity, we use the critical volume fraction of the quark matter in the HQ mixed phase $F_\mathrm{C}$ instead of $\rho_\mathrm{T}$. If the matter has a volume fraction $F > F_\mathrm{C}$ at the density in the layer of the mixed phase, quarks become the color superconducting state.

\begin{figure}[!h]
	\begin{center}
		\includegraphics[width=0.7\linewidth,keepaspectratio,clip]{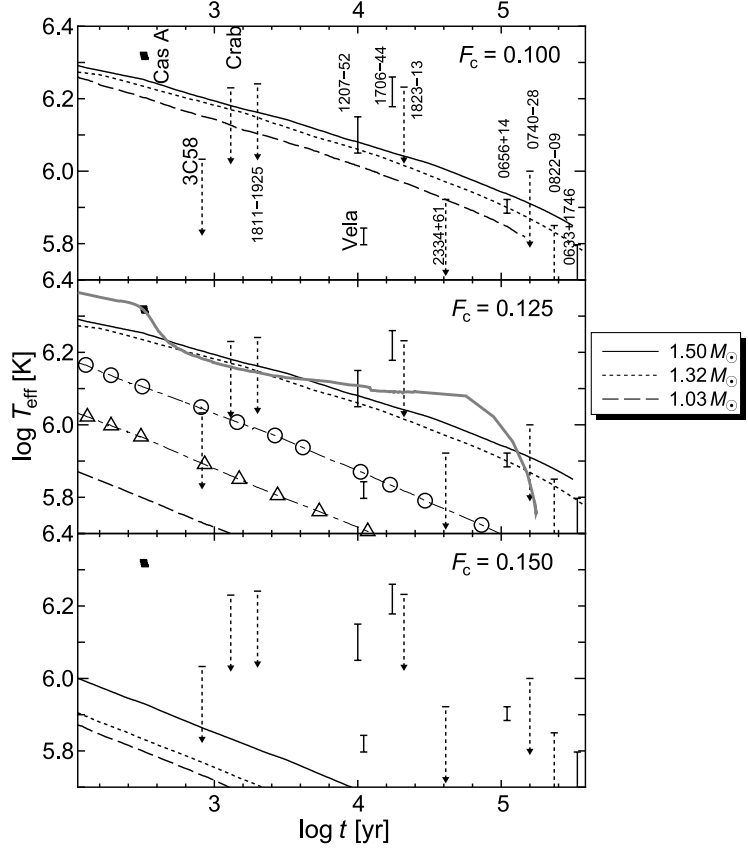}
	\end{center}
	\caption{Cooling curves with color superconducting quark phases~\cite{tn11}. Solid, dotted, and dashed lines denote the models with the masses of $1.50$, $1.32$, and $1.03 ~M_\odot$, respectively. Thick gray line on the middle panel denotes a $1.50M_\odot$ model with PBF and carbon envelope. Dot-dashed lines with marks on the middle panel indicate the model of the mass $1.03 M_\odot$ except for the neutrino emissivity in normal quark phase multiplied by $1/10$ and $1/100$ for the lines with
marks of circle and triangle, respectively.}
	\label{fig:CC}
\end{figure}

We select stellar masses of compact stars to be
$1.0$, $1.3$, and $1.5 M_\odot$ whose central densities are 1.48, 1.82, and
2.67~$\times10^{15}~\rm g~cm^{-3}$, respectively.
We take the critical fraction $F_\mathrm{C}$ to be 0.1, 0.125,
and 0.2
which are appropriate to explain the observations with our scenario.
Results are shown in Fig. \ref{fig:CC},
with available observational values.
Most of the observational data are taken from the Table 7.2 in~\cite{kp06} :~\cite{sl02} for 3C58,~\cite{wk04} for Crab,~\cite{pl01} for Vela,~\cite{bk03} for 0656+14,~\cite{gh02} for 1706-44,~\cite{kp06}
for 1811-1925,~\cite{gs03} for 1823-13,~\cite{bk96} for 2334+61. Other data are taken from~\cite{pg02}. The data of Cas A is attached~\cite{ho10} as the most youngest compact star.
The cooling curves transit from hotter region to cooler region
for the parameter between $0.1 < F_\mathrm{C} < 0.2$. As indicated in the middle panel of Fig. \ref{fig:CC}, cooling curves split into two regions for $F_\mathrm{C}=0.125$ and stars with larger masses cool slower than those with lighter masses. Since lighter mass stars cool faster, they suit for the 3C58 case who does not have lower limit of the effective temperature observation. However, calculated cooling curves are inconsistent with the Vela case which has lower limit and the quark cooling is still too strong to explain this case.

Since the neutrino emissivity of the quark phase involves large uncertainty, we have calculated the additional cooling curves for the mass $1.03 M_\odot$ in case of $F_C=0.125$ with the neutrino emissivity reduced by a factor of $0.1$ and $0.01$.
There are some possible factors of this reduction for neutrino emissivity accompanying quark $\beta$-decay, such as increase of the abundance of strange quarks; decreasing in electron numbers inside MP, leads to the reduction of the neutrino emissivity \cite{iwa80}. The presence of 2SC at low density also suppress the emissivity; \cite{hyp07} discussed that the abundance of quarks in the mixed phase changes and it may cause the color superconducting phase. We suppose that the reduction of the emissivity originates from the above physical processes.
If the emissivity of quarks is reduced by these factors, the observation of Vela can be explained as shown in the middle panel of  Fig. \ref{fig:CC}.
Also, we calculated the effect of PBF and the carbon envelope with our model to fit the recent Cas A data~\cite{ho10}, and the result is shown in the thick gray line in the middle panel of Fig. \ref{fig:CC}.

Considering the HQ mixed phase, the cooling theory of neutron star requires the existence of some kind of color superconducting phase with large gap.
If the mass of an isolated neutron star other than Cas A is detected in near future, it might help to probe the existence of color superconducting phase, quark matter and the HQ mixed phase.

\section{Summary and Concluding Remarks}
\label{sum}

We have discussed features of the HQ mixed phase in compact stars, where multi components are in chemical equilibrium.
Such phase transitions are generally not congruent and the Maxwell construction, which is a common method to obtain the equation of state (EOS) of matter with a single component, cannot be applied any more and one has to solve the Gibbs conditions instead.
Using the bulk calculation, i.e., without considering the surface tension and the Coulomb interaction among two phases,
we have presented basic concepts of the non-congruent transition.
Although these concepts are helpful in discussing and analyzing the properties of the phase transitions
in compact stars, we must bear in mind that they are obtained in an idealized and simplified situation.
For realistic cases, we have to take into account rather complicated finite-size effects such as the surface tension and the Coulomb repulsion.

When the finite-size effects are present,
crystalline structures of regular geometries called ``pasta structures'' emerge in the mixed phase.
Charge screening, i.e., the rearrangement of charged particle densities in the presence of
the Coulomb interaction is also emphasized;
it sometimes causes an instability of the geometrical structures in the mixed phase.
In the extreme case the EOS resembles the one given by the Maxwell construction,
and uniform nuclear matter can exist in the mechanically unstable region.

We note that the maximum mass of hybrid stars in our calculation is rather low
in the light of a recent observation \cite{dem10}, since contamination of hyperons considerably softens EOS.
We have found that the transition of hadronic matter to quark matter suppresses the appearance of hyperons,
but the maximum mass is still low due to the rather soft EOS of quark matter.
Therefore, we need other EOSs of hadron and quark matter to circumvent this situation.

So far, most works on the pasta structures have used the WS cell
with ansatz about the geometrical structures like droplet, rod, slab and so on.
If we want to see the possibility of the intermediate shapes, or study the mixed phase
in a more realistic way,
Frameworks without the WS approximation or ansatz on the geometrical structures are desirable.
Recently we have performed a three-dimensional calculation of non-uniform nuclear matter
based on the relativistic mean-field model and the Thomas-Fermi approximation \cite{oka}.
Introducing a cubic cell with periodic boundary conditions and discretizing it into grids,
we numerically solve the coupled field equations for meson mean-fields and the Coulomb potential.
Randomly distributing fermions ($n, p, e$) on the grid points as an initial condition,
we relax their density distributions to attain the uniformity of the chemical potentials.
We have shown that typical pasta structures, obtained in the previous studies with the WS approximation,
also appear in the three-dimensional calculation~\cite{oka}.

Possibilities of observing the signals of the mixed phase has been suggested
in the spectra of the gravitational waves \cite{SYMT2011}.
We have found that one could know the existence of density discontinuity inside the star via observing the gravitational waves of not only $g$ mode but also $f$ mode.
The detailed observations of $p$ mode gravitational waves could reveal the properties of the HQ mixed phase
because the  $p_1$ mode gravitational waves depend on the adopted EOSs even if the density discontinuity does not exist.

QPO in giant flares  from SGR is another probe from neutron stars.
If the observed QPO frequencies are associated with the torsional oscillations in neutron star crust, one might be able to obtain the information about the crust properties.
In fact, we have pointed out that some of the QPO frequencies observed in giant flares might be associated with the torsional oscillations in the HQ mixed phase.

Phenomenological implications of pasta structures should deserve more elaborate studies of
their thermodynamical or dynamical properties, e.g.,
the neutrino opacity or the thermal conductivity may be affected by the pasta structures,
and their elasticity may cause a discontinuous change of internal structure of compact stars.
In particular,
the crystalline structure of the HQ mixed phase in the core region may bring about
novel dynamical effects on compact-star phenomena
by way of deformation or oscillation.

The other important application of the EOS with the HQ mixed phase is the cooling theory of neutron stars, the long time issue discussed for decades.
The cooling of neutron stars strongly depends on the neutrino emissivity of the core of stars.
Simply applying the HQ mixed phase with normal quark matter, stars cool rapidly and do not match observations.
However, including color superconductivity, the neutrino emissivity of the quark phase is much suppressed, and the cooling of stars with the HQ mixed phase does not conflict with observations.

Taking account of a color superconductivity in the HQ mixed phase, it became possible to have heavier stars cooling slower and lighter ones faster.
We cannot get this situation without considering the HQ mixed phase.
If one observes the masses of isolated neutron stars and the mass--temperature distribution can be described by the above scenario, it proves the existence of the color superconductivity in the HQ mixed phase.

\section*{Acknowledgements}
NY is grateful to H.-J. Schulze, F. G. Burgio, and M. Baldo for their warm hospitality and fruitful discussions.
This work was partially supported by the
Grant-in-Aid for the Global COE Program
``The Next Generation of Physics, Spun from Universality and Emergence''
and for Scientific Research on Innovative Areas
from the Ministry of Education, Culture, Sports, Science and Technology
(MEXT) of Japan  and the
Grant-in-Aid for Scientific Research (C)
and for Young Scientists (B) from JSPS (16540246, 20540267, 23105711, 24740177).

\printindex
\end{document}